\documentclass[aps,onecolumn,preprint,superscriptaddress,nofootinbib,floats]{revtex4}
\usepackage{amsmath,amssymb,color,mathrsfs, graphicx,verbatim,epsfig, bbm, wasysym, multirow}

\def\gaugino{\tilde{\chi}}
\def\stau{\tilde{\tau}}

\def\beq{\begin{equation}}
\def\eeq{\end{equation}}
\def\beqarray{\begin{eqnarray}}
\def\eeqarray{\end{eqnarray}}
\def\ttbar{$t\overline{t}$ }

\def\bit{\begin{itemize}}
\def\eit{\end{itemize}}
\def\sq{\tilde{q}}
\def\bino{\tilde{B}}
\def\wino{\tilde{W}}
\def\gluino{\tilde{g}}
\def\slep{\tilde{l}}
\def\snu{\tilde{\nu}}
\def\sel{\tilde{e}_R}
\def\squark{\tilde{q}}

\def\met{$\displaystyle{\not}E_T$}
\def\tr{{\rm Tr}}

\def\simgt{\mathrel{\lower2.5pt\vbox{\lineskip=0pt\baselineskip=0pt
           \hbox{$>$}\hbox{$\sim$}}}}
\def\simlt{\mathrel{\lower2.5pt\vbox{\lineskip=0pt\baselineskip=0pt
           \hbox{$<$}\hbox{$\sim$}}}}

\begin{document}
\begin{flushright}UMD-PP-09-061
\end{flushright}
\vspace{0.2cm}

\title{Signals of a Sneutrino (N)LSP at the LHC}

\author{Andrey Katz}
\email{andrey(at)umd.edu}
\affiliation{Department of Physics, University of Maryland,\\
College Park, MD 20742}

\author{Brock Tweedie}
\email{brock(at)pha.jhu.edu}
\affiliation{Department of Physics and Astronomy, Johns Hopkins University,\\
Baltimore, MD 21218}

\date{\today}

\begin{abstract} 
\noindent The sneutrino is a viable candidate for the
NLSP in SUSY spectra with gravitino LSP. In this 
work we study the collider implications of this possibility. In particular, we 
investigate whether the
LHC can distinguish it (at least, in some cases) from alternative spectra,
such as those with a neutralino LSP.
We show that there exists a complete family of experimentally allowed and theoretically
motivated spectra with sneutrino NLSP, which exhibit very distinctive multilepton signals
that are difficult to fake within the MSSM.
We study these signals in detail, including the techniques necessary to find them. 
We demonstrate our analysis approach on simulations incorporating backgrounds.
\end{abstract}

\maketitle

\section{Introduction}
\label{intro}

Supersymmetry (SUSY) has been a leading framework for physics beyond
the Standard Model (SM) for decades.  While the
basic motivation is quite simple -- to stabilize the electroweak
sector against quantum corrections from arbitrarily high energy scales 
-- SUSY has proved to be a seemingly never-ending
source of investigation from the perspective of
formal theory, model building, and phenomenology.
The vast parameter space of even the simplest incarnation, the
Minimal Supersymmetric Standard Model (MSSM), leads to a large
variety of possible collider and cosmological
signals, many of which are not yet fully understood.
In this paper, we turn our attention to some of the
less-explored regions of the MSSM parameter space,
regions where the sneutrino appears to be the lightest
supersymmetric particle (LSP) from the perspective of colliders.

There are a couple of reasons why the sneutrino is typically
overlooked as a possible LSP.  In particular, it is
unworkable as a dark matter candidate, since it would have
already been discovered in direct detection experiments~\cite{Falk:1994es,Arina:2007tm}.
There are of course many
ways to model-build around the direct detection constraints, but by
far the simplest is to assume that the sneutrino
decays, either through $R$-parity-violating interactions or
into a lighter gravitino.  The role of dark matter must
then be played either by the gravitino or by some other sector
of particles.  Putting the sneutrino near the
bottom of the spectrum therefore appears to
offer no ``benefit.''  However, even the standard scenario with a neutralino LSP
is becoming
progressively less viable (see, e.g.,~\cite{ArkaniHamed:2006mb}), and we
should definitely not take for granted that it is correct.

Cosmology aside, there is also a simple parametric bias that usually disfavors light sneutrinos.  
In both gravity mediation and gauge mediation 
(GMSB)~\cite{Dine:1981gu,Dine:1982zb,AlvarezGaume:1981wy,Dine:1994vc,Dine:1995ag}, 
the masses of the sfermions acquire
contributions proportional to gauge couplings.  In more minimal scenarios, this
tends to make the left-handed (LH) slepton doublet heavier than the right-handed
(RH) slepton, as well as the bino and often the wino.  However, as we will see in the
next section, it is 
straightforward for the slepton doublet to end up much lighter when we
go beyond minimal assumptions.

Though not often considered, then, there is 
no reason why the sneutrino cannot be at the bottom of the MSSM spectrum.
Nonetheless, the most distinctive
collider implications of this possibility remain mostly
uninvestigated. 
If we assume that $R$-parity holds as a good
symmetry, which is still very well-motivated from the perspective
of forbidding proton decay, then the minimal cosmologically-viable
scenario has the would-be LSP sneutrino decaying into the gravitino.
However, since this decay is completely invisible, the sneutrino
acts like an LSP in colliders, independent of its lifetime.\footnote{Our analysis
will also trivially extend to those cases with $R$-parity violation where the
sneutrino decays outside the detector.}

Here, we will begin an inquiry into the LHC signals of this sneutrino
``(N)LSP''.  We will find decay chain topologies and collider signatures
that are noticeably different from the standard cases, such as a neutralino LSP.

Of course, even assuming a sneutrino NLSP with gravitino LSP, the parameter space describing
the rest of the spectrum remains enormous.  In our analysis here, we will concentrate on 
spectra with $\mathcal{O}$(TeV) colored superparticles and lighter electroweak gauginos,
all sitting above approximately flavor-degenerate LH slepton doublets. However, many of 
our observations will also survive in more generic spectra.

Given this setup, we can identify two broad classes of spectra which lead to somewhat
different phenomenology:  either the RH slepton participates in the decay chains,
or it mostly does not.  Roughly speaking, this depends on whether the RH slepton is 
heavier or lighter than the bino-like neutralino.  In this paper, we will concentrate
on the collider signatures of the simpler case where $m(\sel)> m(\bino)$.  
We will investigate spectra
with active RH sleptons in a companion paper~\cite{active}.

In every decay chain ending in a sneutrino NLSP, lepton number is being carried
away.  This must necessarily be compensated for by
release of a
charged lepton or a neutrino.  In addition, the sneutrino is
quasi-degenerate with its $SU(2)_L$ partner, the charged LH slepton,
so that the two sfermions will usually sit together at the bottom of the spectrum.  This 
means that every decay chain has a sizable chance of 
producing at least one charged
lepton, significantly modifying the lepton accounting compared to
many alternative spectra.

In addition, the splitting within the
slepton doublet is often large enough to produce visible particles
via $W^*$ emission.  In the leptonic mode, this can come in
combination with a charged lepton produced with the slepton, leading to a
rather unique excess of opposite-sign uncorrelated-flavor
lepton pairs.  This excess
will be accompanied by somewhat larger number of completely sign- and 
flavor-uncorrelated dilepton events, as well as a somewhat smaller number 
of very distinctive trilepton events.  The coexistence of all of these features
together is highly non-generic in the MSSM, and will serve as a powerful
indicator of the presence of a sneutrino NLSP.

Our paper is organized as follows. 
In the next section, we give  top down motivations for this scenario,
discussing classes of mediation models in which
a sneutrino NLSP may arise.  
In section~\ref{genpr}, we summarize the details of 
the slepton/sneutrino spectrum, decays, and cosmology.
In section~\ref{cascades}, we discuss generic event topologies
and lepton accounting.  (Readers mainly interested in collider phenomenology can
start here.)  We then perform a more detailed study of LHC signals
in section~\ref{signat}, including estimates of the dominant backgrounds.
Section~\ref{conc} finishes with some closing thoughts and ideas for
further studies.

\section{How to get a sneutrino NLSP}
\label{motiv}

\subsection{Gauge Mediation}

Giving up on a cosmological role for a light sneutrino, it finds a natural
home in gauge mediation, where all would-be LSP's are fated to decay into the
gravitino.  Indeed, a sneutrino 
NLSP readily
arises is General Gauge Mediation~\cite{Meade:2008wd}.
GGM is defined as the class of UV completions
of the
MSSM which fulfill the following condition: in the limit where all gauge
couplings are taken to zero, the theory decouples into the SUSY breaking sector
and the MSSM with exact SUSY. This definition includes various
perturbative
messenger models (either of direct gauge mediation or including a separate
messenger sector), as well as nonperturbative ones, where even the
definition of the messenger field is obscure. In spite of the fact that
this definition covers a very broad class of models with potentially very
different collider signatures, all of these models possess some common features,
like maintaining the sum rules $\tr\left[Ym^2\right] = \tr\left[(B-L)m^2\right] = 0$ 
for the sfermion soft masses 
(at the messenger scale), as well as parametrically suppressed $A$-terms.

GGM can be fully parameterized by six independent
parameters at the messenger scale (not including the Higgs 
sector\footnote{One needs an additional mechanism (which might not
fit the definitions of GGM) in order to produce reliable $\mu$ and $B\mu$
terms. Such mechanisms usually significantly modify the soft masses of the
Higgses from the GGM value (see e.g.~\cite{Csaki:2008sr,Komargodski:2008ax}).
Therefore, in exploring the parameter space of GGM one should keep the
Higgs sector scales as free parameters.}
and the 
value of the messenger scale itself). 
One can think about this parametrization as follows: each gauge
group of the standard model (SM) has some contribution to its gaugino and
an independent contribution to the scalars charged under that group. We can
parameterize the soft masses as
\beq
M_r = g_r^2 M B_r, \ \ \ m_{\tilde f}^2= \sum_{r=1}^3 g_r^4 C_2( f, r)
M^2 A_r~.
\eeq
In these formulas $r$ runs over the three gauge
groups of the SM.  $B_r$ are three complex numbers, and $A_r$ are three real
numbers. The $g_r$ are the gauge couplings, and $C_2( f, r)$ denotes the quadratic 
Casimir of $f$ with respect to the
gauge group $r$. $M$ is an overall soft mass scale.

The GGM framework admits a large variety of different spectra, 
far beyond that of  minimal GMSB models. Not all of these spectra are 
fully understood in terms of  collider signatures and experimental 
constraints. For example, the signatures of the promptly decaying 
neutralino NLSP at the  Tevatron have been only recently studied in full generality
in~\cite{netralinotev}. The chargino, which can also be the NLSP in a narrow 
region of GGM parameter space, has also been neglected for a long time
and was first seriously considered only recently~\cite{Kribs:2008hq}. 
   
Obtaining sneutrino as the NLSP was found to be straightforward 
in~\cite{Carpenter:2008he,Rajaraman:2009ga},
even with restrictions on the GGM parameters.  
For example,
in order to ensure that the LH slepton doublet is lighter than the RH slepton, 
it is adequate simply to demand\footnote{Note that this is only a rough upper
bound on $A_2$, since it does not account
for radiative corrections or left/right mixing effects.  In particular, near the
boundary, stau mixing
effects can become large, and $\tilde{\tau}_1$ may become the NLSP.}
\beq
A_2 \simlt \frac35 \frac{g_1^4}{g_2^4}A_1 \simeq (0.2)A_1~.\label{eq:A2}
\eeq
The 
gauginos can subsequently be made heavier than the LH slepton completely 
independently by adjusting the $B_r$, and the squarks can easily be made heavier 
as well. 

Realizing this situation with perturbative models is also possible, 
though not quite so trivial.  Using the GGM notation, the $B$s are now
related to the $A$s.  Specifically,
in order to make the wino arbitrarily heavy with respect to the LH slepton
in models with purely $F$-term SUSY-breaking masses for the messengers,
we would need arbitrarily large messenger Dynkin index.\footnote{The messenger Dynkin
sets upper bounds on the mass ratios between the gauginos and the sfermions.
Indeed, it is straightforward to engineer models that reduce these ratios, 
but we know of no examples of the opposite effect.}
However, perturbative gauge unification
is spoiled unless the Dynkin is $\simlt 5$, or the messenger scale is made very 
high.  Taking the former constraint, the best we can do is to make the wino
a little less than twice as heavy.  At the same time, we must have a large
mass for the RH slepton (which automatically also translates into a large
mass for the bino), and this also feeds into the LH slepton via $A_1$, decreasing
the gap with the wino.  Without pushing to large Dynkin,
the best we can manage is a rather squashed
spectrum of sleptons and electroweak gauginos at the messenger scale, and 
mixing effects and radiative corrections can easily reorder it.  

A simple model that manages to produce an acceptable spectrum contains
two identical copies of $\bf{10}$+$\overline{\bf{10}}$, with independent
supersymmetric and supersymmetry-breaking masses for each
of the SM irreducible representations.\footnote{This 
can easily be accomplished by coupling in multiple singlet fields which 
feel SUSY breaking, assuming the couplings are not $SU(5)$ 
symmetric~\cite{Carpenter:2008wi}. However,
since the Dynkin index here is 6, the messenger scale must be 
$\mathcal{O}(1000)$ TeV or larger to permit perturbative gauge 
coupling unification.  In a
more complete analysis, one also needs to worry about threshold 
corrections from the explicit global $SU(5)$-breaking
in the messenger spectrum.}
From the perspective of the electroweak gauginos and sleptons, the $Q$-like
($(\bf{3},\bf{2})_{1/6}$ + cc) messenger  fields act 
as an approximately pure source of 
$A_2$ and $B_2$, 
and the $U$-like ($(\bf{3},\bf{1})_{-2/3}$ + cc) and $E$-like ($(\bf{1},\bf{1})_{1}$ + cc) 
fields act as independently tunable sources of $A_1$
and $B_1$.
The LH slepton is lightest by a comfortable margin when we
choose parameters such that
$B_1$ is $2 \sim 3$ times larger than $B_2$.

More generally, we can consider perturbative models 
with $D$-term contributions from a hidden gauge sector under which the messengers
are charged.  This was suggested in~\cite{Buican:2008ws} as a way 
to fully realize GGM without resorting to strong coupling.
In the case at hand, it allows us to independently decrease the LH slepton mass 
with respect to the wino mass without requiring
arbitrarily large Dynkin index.

Clearly, then, a sneutrino NLSP in gauge mediation would point to a highly 
non-minimal, possibly nonperturbative messenger sector.  Of course, even if 
the sneutrino NLSP is actually established at the LHC, determining whether
the spectrum is due to non-minimal gauge mediation will be quite difficult, 
since even a prompt decay of the sneutrino is completely invisible.  Still, 
there will be various clues encoded in the spectrum, namely the sum rules above, near
flavor-universality (and small $A$-terms), and perhaps even the apparent 
high-scale non-unification of gaugino masses.

\subsection{Other Scenarios}

Obtaining the sneutrino at the bottom of the superparticle spectrum is also possible
with other mediation mechanisms, such as gravity mediation or gaugino
mediation~\cite{Kaplan:1999ac,Chacko:1999mi}.\footnote{In spite of the fact that the
gravitino is not a guaranteed LSP candidate in these scenarios, it is a logical possibility
which we further consider.}  

One way to accomplish this~\cite{Ellis:2002iu} (see also~\cite{Buchmuller:2005ma, Evans:2006sj})
is to invoke highly non-universal 
Higgs masses (NUHM) at the input scale.
This leads to large hypercharge $D$-term 
loop contributions in the renormalization
group equations, what is usually called the ``$S$-term''.  It is defined as
\beq
S \equiv \tr\left[ Y m^2 \right],
\eeq
where the trace runs over all soft scalar masses of the MSSM, including the Higgses.
 It contributes
to the running of the superpartners as 
\beq
\Delta\left(\frac{d m^2}{d\ln \mu}\right) = \frac{1}{8\pi^2}\left(\frac35 g_1^2 Y\right) S~.
\eeq
In parameterizations of high-scale mediation with universal soft masses, this 
term vanishes at the scale where the soft masses are produced.  However, by 
taking the Higgs mass parameters independent of the sfermions, and making the 
down-type mass much larger than the up-type, the $S$-term can push the RH 
slepton mass higher than the LH slepton mass.

Since the down-type Higgs mass is very large in these scenarios, the contributions 
to the running of the third generation masses is significant.  In particular, the 
third-generation slepton doublet can be appreciably lighter than the other two, at 
the level of 10's of GeV.  This has led to several 
discussions~\cite{Covi:2007xj,Ellis:2008as,Medina:2009ey,Santoso:2009qa} 
that emphasize signals with taus.  We will not pursue this approach here,
but note that large flavor non-degeneracies within the sneutrinos can lead to
additional variations on the more general signals that we will explore in 
section~~\ref{cascades}.  
We defer investigation of these to~\cite{active}.  

Another, simple way to obtain a sneutrino NLSP in gravity mediation is 
by using independent (but flavor-universal) masses-squared
for the {\bf 5} and {\bf 10} representations of $SU(5)$.  The RH slepton
can be made arbitrarily heavier than the LH slepton in this way, but 
in order to end up with the LH slepton lighter than the bino, 
we should go to negative $m_{\bf 5}^2$.
The tachyonic masses for the LH slepton and RH down squark will 
run positive in the IR, provided they are
not too large in magnitude, and they should not pose
any phenomenological difficulties.\footnote{Universal negative masses-squared
were considered in~\cite{Feng:2005ba}.}  Since the Higgs soft masses are not 
necessarily very large, flavor non-universal contributions to the sleptons will 
typically be smaller.  These spectra should fall under our analysis 
here.

Physics beyond the MSSM could also play a role.  For example, the LH slepton
can be rendered light via renormalization group running induced by
couplings to right-handed neutrinos, as recently proposed in~\cite{Kadota:2009sf}.
Even though this paper did not consider spectra with sneutrino NLSP and
\emph{gravitino} LSP, it is likely possible to achieve with the same
kind of setup.
Since the physical mass spectrum here depends on new, unknown Yukawa couplings,
flavor violation effects may be arbitrary.  This is an interesting possibility, but
falls outside the scope of our present work.

In this paper, we will consider spectra that are relatively flavor-univeral.  We
have noted that a gravity mediation scenario with negative $m_{\bf 5}^2$ may fall
into this class, but most of the alternatives involve some non-negligible degree of flavor violation,
for example leading to tau-enriched signatures.
In particular, the discussions 
of~\cite{Covi:2007xj,Ellis:2008as,Medina:2009ey,Santoso:2009qa,Kadota:2009sf} are
essentially independent of our own observations below regarding flavor-universal multi-lepton signals.

\section{General properties of models with a sneutrino NLSP}
\label{genpr}

In this section, we discuss in more detail the generic features of the
LH slepton and sneutrino states.  We first work out the fine-structure
of the mass spectrum, then discuss decays of the charged sleptons, 
and finally consider the role of the sneutrino 
NLSP in cosmology.

\subsection{Slepton and Sneutrino Mass Spectrum}

Before electroweak symmetry breaking, all three flavors of
sneutrino are precisely 
degenerate with their 
charged partners.  The measured difference between mass eigenstates is 
dictated by electroweak $D$-terms 
and by the mixing between the left- and right-handed sleptons.

The $D$-terms act to make the charged sleptons heavier than the
sneutrinos for $\tan\beta > 1$:
\beq
m_{\slep_L^+} - m_{\snu} = \frac{m_W^2 (-\cos(2\beta))}{m_{\slep_L^+} 
+ m_{\snu}} \simeq \frac{m_W^2 \sin^2 \beta}{m_{\slep_L^+} + m_{\snu}}~,  
\label{eq:hyperfine}
\eeq
where in the last term we have displayed the large $\tan\beta$ limit, 
which becomes accurate
to better than ${\mathcal O}(10\%)$ for $\tan\beta \simgt 3$.  The splitting
is inversely proportional to the average doublet mass, and is always
less than $m_W$.  For example, for masses near 200 GeV, the 
splitting is about 16 GeV or smaller.

In the third generation, left/right mixing effects might become important.  
The mass-squared matrix of the $\stau$s is
\beq
\left(\begin{array}{cc}
 m^2_{\stau_L}+\Delta_L & -\mu v y_\tau \sin \beta+vA_\tau^*\cos \beta \\
-\mu^* v y_\tau \sin \beta +vA_\tau\cos \beta  & m^2_{\stau_R} +\Delta_R
\end{array}
 \right)~,
\eeq
where we have represented the electroweak $D$-term 
contributions to
LH and RH sleptons as $\Delta_L$ and $\Delta_R$, respectively.
If we neglect $A$-terms, and if the the $\mu$/Yukawa terms 
are not too large compared to the mass-squared difference,
the (mostly) left-handed stau is shifted down by approximately
\beq
 - \frac{\mu^2 v^2 y_\tau^2\sin^2 \beta}{m_{\stau_R}^2-m_{\stau_L}^2}~.
\eeq
(The $D$-term contributions have been neglected in the denominator,
but they are small, roughly $(20 \; {\rm GeV})^2$.)
This can potentially make $\stau_1$ the NLSP, if the mass correction from mixing 
is larger than the electroweak doublet splitting above.  Clearly then, in any 
viable sneutrino NLSP scenario, we will require\footnote{The LH stau doublet 
also receives flavor-nonuniversal mass corrections at 
loop level.  These tend to push the third generation lighter, making the 
tau-sneutrino lightest.  If these corrections are large, such as in
high-scale scenarios with large Higgs soft masses, then $\stau_L$ can 
be lighter
than the sneutrinos of the first two generations, even before left/right mixing
is taken into account.  We will not consider such scenarios here, but see~\cite{active}.}

\beq
\frac{\mu^2 v^2 y_\tau^2}{m_{\stau_R}^2-m_{\stau_L}^2} \simlt m_W^2~.
\eeq

In general then, spectra with sneutrino NLSP can have quite
detailed fine-structure for the lowest-lying states.  Here, we will 
be interested in cases with approximate flavor degeneracy between
the three slepton doublets, which carries over to the three flavors of 
NLSP sneutrinos.  The stau-sneutrino will be the true NLSP, but the 
splitting with the other sneutrinos will not be large enough to generate
any visible activity.  Stau mixing will always push the
$\stau_1$ lighter than the first- and second-generation $\slep_L^+$, 
but as long as it stays heavier than the sneutrinos, the modifications
to the phenomenology we consider here are minor.

\subsection{Decays}\label{decays}

\begin{figure}[t]
\centering
\includegraphics[width=6.7in]{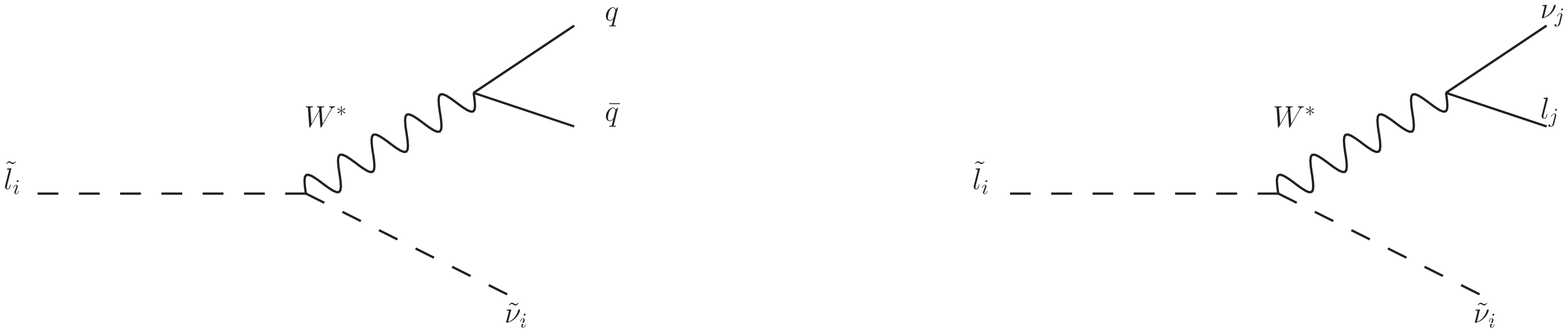}
\caption{Possible decay modes of the LH slepton. While the first diagram 
produces low-multiplicity jets, the second produces a relatively soft lepton.
This lepton can be visible above backgrounds, if accompanied by other hard activity.}
\label{fig:1}
\end{figure}

In equation~\eqref{eq:hyperfine}, we saw that in general the mass splitting 
between the sneutrino and its charged slepton partner is always less than $m_W$.  
If all of the charginos and neutralinos are heavier, then the charged slepton
decays will be dominated by $W^*$ emission:  $\slep \to \snu f \bar f'$ 
(see Fig.~\ref{fig:1}).\footnote{Virtual gauginos will also contribute to the
decays, opening up additional modes, and interfering with some of the $W^*$
modes in a flavor-dependent way.  Practically, these tend to be much less important
if the gauginos are well above $m_W$.}  
(Subsequently, $\slep$ will always refer to $\slep_L$, so we drop
the subscript.)
As long as 
$m_{\slep^+} - m_{\snu} \simgt$ few GeV (which for even modest $\tan\beta$ is 
almost always true), 
the branching fractions into different
species of quarks and leptons will be very similar to that of on-shell $W$s.
In particular, it will produce $e$ or $\mu$ approximately
22\% of the time.  It can also produce leptons from secondary $\tau$ decays,
but the vast majority of the other decay modes will contain low-multiplicity jets.
These will be quite difficult to cleanly identify, and we will not explore 
the possibility of using them in our searches.

Seeing the products of the $e$ and $\mu$ decay modes could in principle be 
complicated by the
fact that the slepton/sneutrino mass splitting can become quite small, and 
that it is three-body.  In the example given in the previous subsection, 
of a 200 GeV doublet split by 16 GeV,
the average lepton momentum in the slepton rest frame is about 8 GeV.  This 
approaches the threshold for good quality lepton identification in the LHC experiments. 
So a fraction of the leptons, from the softer region of the emission spectrum, 
may be unobservable.
Since the hardness of the emitted leptons scales inversely with the doublet mass, 
heavier sleptons will yield softer leptons, possibly leading to ${\mathcal O}(1)$
loss of signal.  This may be partially compensated for in strong SUSY production, 
in which the sleptons may be produced with substantial boost from decay chains
initiated by much heavier squarks and gluinos.

Since for cosmological reasons, we work with models where the 
sneutrino is the NLSP and the gravitino the LSP, we should also be mindful
of the possibility that the charged LH slepton can decay directly to the 
gravitino and a charged lepton, much like a RH slepton NLSP.
Though suppressed by some mass scale significantly larger than $m_W$, this decay 
is two-body, and typically has much more phase space than the $\snu l \bar \nu$ mode.

Assuming $m_{3/2} \ll m_{\slep}$, the decay width into gravitino and 
lepton is given by 
\beq
\Gamma(\tilde l \to \tilde G l) \simeq \frac{m_{\slep}^5}{16 \pi F^2}~, \label{gravdecay}
\eeq   
where $\sqrt{F}$ is the fundamental SUSY-breaking scale.  The decay
width into $\snu l \bar \nu$ by $W^*$ emission goes as
\beq
\Gamma(\slep \to \snu l \bar \nu) \simeq \frac{\alpha_2^2}{30 \pi}
\frac{(m_{\slep^+}-m_{\snu})^5}{m_W^4} \simeq \frac{\alpha_2^2\,|\cos(2\beta)|^5}{2^6\cdot 15 \pi}
\frac{m_W^6}{m_{\slep}^5}  ~,
\eeq 
where have used equation~\eqref{eq:hyperfine} and taken the small splitting limit.  
The two-body decays become competitive if
\beq
\sqrt{F} \simlt \left(\frac{60}{\alpha_2^2\,|\cos(2\beta)|^5} \frac{m_{\slep}^{10}}{m_W^6}\right)^{1/4} \simeq
\; (2 \; {\rm TeV})  \; \frac{1}{|\cos(2\beta)|^5}\left(\frac{m_{\slep}}{100\;{\rm GeV}}\right)^{5/2}~.
\eeq
For modest values of $m_{\slep}$, such a small value of $\sqrt F$ is difficult to obtain in known models.
For bigger values of  $m_{\slep}$, the two-body decays will be relevant for a small part of the viable 
parameter space.  In subsequent sections, we will simply assume that the 
SUSY-breaking scale is not so small, and that two-body decays are subdominant.  
However, cases with significant two-body contributions would also be interesting
to study.\footnote{If two-body dominates, then the phenomenology would look very 
similar to RH slepton NLSP, but with
modified lepton counting.  Instead of every event containing at least two hard 
leptons (and/or taus), each decay chain would now have a $\mathcal{O}(50\%)$ chance
of ending via $\snu \to \tilde G \nu$.  In a scenario with mixed two-body 
and three-body decay events, the former will introduce an
additional population of opposite-sign same-flavor (OSSF) leptons from charged
slepton production and decay.}

\subsection{Cosmology}

A sneutrino NLSP decays mostly invisibly, into neutrino and gravitino,
and its would-be relic density becomes attenuated by $m_{3/2}/m_{\snu}$.
For most choices of parameters, it decays before matter domination.  
Therefore the neutrinos
generated in its decay would presently constitute a subdominant non-thermal 
component of the cosmic neutrino background.  In the cases where the gravitino
is heavy enough to deposit a visible amount of kinetic energy in terrestrial 
nuclear recoil experiments, the gravitino-on-nucleon cross section is suppressed 
to a point well below present bounds.

The possibility that the gravitino produced in the decay is in fact dark matter,
or some component of it, has been considered in a number of papers (see, for example,
\cite{Bolz:2000fu,Feng:2003xh}).  Since
we are mostly interested in collider signals, we will not explicitly consider
this.

Despite the elusive nature of the sneutrino's main decay products, rare
decays involving on-shell or off-shell gauge bosons do produce visible particles.
If the lifetime of the sneutrino is longer than a few seconds, these can
upset the light element abundance of nucleosynthesis.  This possibility
was considered in detail in~\cite{Kanzaki:2006hm}.  For sneutrinos lighter than about
300 GeV, and/or gravitinos lighter than about 4 GeV, there is no constraint.
In particular, low-scale mediation scenarios like gauge mediation are 
automatically cosmologically safe.

\section{SUSY cascades with a sneutrino NLSP}
\label{cascades}

When all decay chains end in the LH slepton doublet, there can be significant 
modifications compared to more standard MSSM spectra.  In this section, we 
will discuss these features and categorize promising decay topologies for 
searches at the LHC.

The simplest SUSY events available 
at the LHC are direct pair production of the sneutrino and/or 
its charged partner, accompanied by a decays through $W^*$ in the latter 
case (Fig.~\ref{fig:1}).  In a hadron collider environment, none of these 
options is 
particularly easy to find.  Sneutrino pair production is completely invisible, 
and could only be detected in principle through ISR, essentially impossible 
at the LHC.  Including the charged slepton leads to relatively soft leptons 
and/or jets with very little missing energy.  It will likely be swamped by 
physics backgrounds and fakes, even in the dileptonic mode.

We are therefore led to consider more complicated processes, where the sneutrino
is the final particle emitted in cascades initiated by gauginos or squarks/gluinos.
As usual, the most striking situation is $\mathcal{O}$(TeV)
super-QCD (SQCD) production, and we choose
to focus on this case.\footnote{Squarks near 1 TeV are actually favored
to achieve $m_H > 114$ GeV (via loop corrections) in the MSSM, in the absence 
of large $A$-terms and with near flavor degeneracy.  Of course, it is not 
difficult to circumvent this in extensions of the 
MSSM~\cite{Dine:2007xi}, for example~\cite{Nomura:2005rk,Nomura:2005rj}.
Similar issues were earlier discussed in~\cite{Brignole:2003cm,Casas:2003jx}.}
Electroweak (EW) production of gauginos may also be visible,
particularly in trilepton (or higher) channels.
These become particularly important when the squarks and gluinos are much heavier
than a TeV,
and therefore difficult to produce.  We do not study these explicitly, but 
the general observations below for the structure of trilepton signals in 
SQCD production should carry over essentially unchanged.

Before we proceed, let us summarize our main assumptions,
and describe more carefully the types of spectra we will analyze: 
\bit

\item The RH sleptons are heavier than the (mostly-)bino,
\beq
m(\sel)> m(\bino)~.
\eeq
This is sufficient to ensure that the RH slepton is bypassed
in the vast majority of the decay chains.  We will consider spectra with
this inequality reversed in~\cite{active}.

\item Approximate flavor degeneracy.  In particular, the fine structure
of the three LH slepton doublets is dominated by the $D$-term splitting
in equation~\eqref{eq:hyperfine}.  The sneutrinos are mass-degenerate
for all practical purposes.  The mostly-LH stau may be somewhat lighter
than the other charged sleptons, but still heavier than the sneutrinos.

\item All other superparticles are heavier than the charged LH slepton states.
 
\item $\mathcal{O}$(TeV) colored superpartners, sitting above the EW gauginos.

\item Modest gaugino-Higgsino mixing.

\item Negligible $A$-terms at the mediation scale.

\eit 

In the following subsections, we will explore the signals of this class
of scenarios in more detail, as well as the difficulty of replicating them
in alternative MSSM spectra.  The discussion is at the conceptual level,
without worrying about many of the complications of the hadron collider
environment.  We will perform a more realistic analysis illustrating these
ideas in simulation in section~\ref{signat}.

\subsection{Lepton Counting}\label{count}

Consider, as a first stage, simple counting of the leptonic events including  
flavor/sign correlation between the leptons in dileptonic events.  We will
see that this lepton accounting is quite distinctive with a sneutrino NLSP.

Starting from squark/gluino production, each decay chain must proceed down
to the slepton doublet through charginos and neutralinos.
Since we will be assuming that the colored superparticles are heaviest, these
will always be treated as on-shell.  However, this point 
is not crucial for many of our observations.

We also neglect decay 
chains containing (mostly-)Higgsinos.  These will be
relevant for chains initiated by stops and sbottoms, but they will not typically
serve as more than a correction to our signals.\footnote{Spectra with strongly
mixed charginos and neutralinos are an obvious exception to this.  Also, in relatively
unmixed cases with very light Higgsinos, they may contribute to the 
EW gaugino decay chains, via Higgs or electroweak boson emission.}

\begin{figure}[t]
\centering
\includegraphics[width=6.7in]{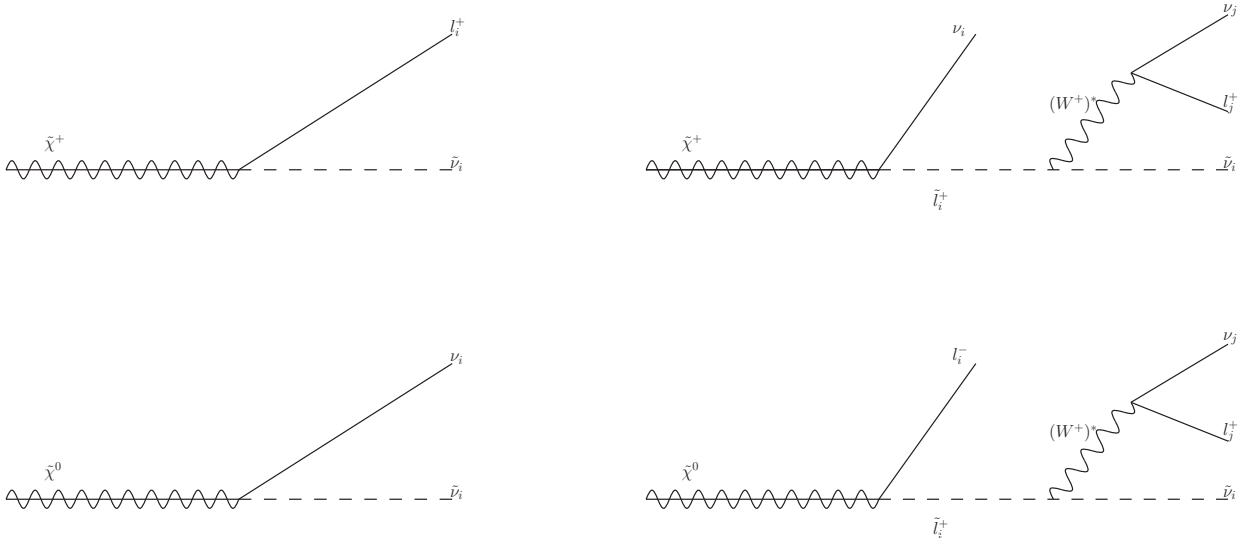}
\caption{Possible decay modes of gauginos, with $W^*$
decaying leptonically.  Notice 
the sign and flavor flow on the last diagram, with two charged leptons in a single decay chain.
While the leptons' signs are correlated, their flavors are not.}
\label{fig:2}
\end{figure}

We therefore concentrate on the decays of charginos and neutralinos that are
mostly wino and bino.  We illustrate the decay chains in Fig.~\ref{fig:2}.
Up to possible phase space suppressions, and coupling shifts due to gaugino
mixing, \emph{any} 
EW gaugino decay has a 50\% probability of producing a charged lepton in the first stage
of the decay chain.
Since the decays are flavor-blind, 1/3 of these leptons will be taus.  At first
pass we will treat these as ``hadrons,'' considering only the prompt production
of electrons and muons.  This still leaves 1/3 of the gaugino decays, whether chargino
or neutralino, going into one of these highly visible lepton modes.  This leads to a 
significant chance for each SUSY event to contain one or two hard, isolated leptons.
Specifically, the ratio of dilepton:monolepton:no-lepton would be roughly 1:4:4.

If this had been  the only source of the isolated leptons, 
then the dilepton signal would have been completely charge 
and flavor uncorrelated, since each lepton is produced in a different 
chain.\footnote{Sign correlation across the event \emph{can} occur to some extent.
For example, we can have LH squark-antisquark pair production followed by decays into
charged winos on both sides, each subsequently decaying into charged leptons.  This
leads to opposite-sign leptons.
Similarly, squark-squark production (from valence quarks annihilating via gluino exchange) 
may have a bias towards same-sign leptons.  Practically, neither of these tends to lead to an 
overwhelming charge bias, when all production and decay modes are taken into account.}
However, we also have leptons from the charged slepton's decay via $W^*$.  Although these
leptons tend to be softer, due to the approximate slepton-sneutrino degeneracy,
they are still often visible.  To some extent, these add to the monoleptonic 
signal, but their effects at higher multiplicity are much more interesting.

In the dileptonic events, we now have contributions where both leptons
come from a single neutralino decay chain (\emph{either} bino or neutral wino), 
as in the lower-right diagram of Fig.~\ref{fig:2}.  Since the probability
of the first decay to produce an electron or muon (plus selectron or smuon)
is 1/3, and the probability
of the second to do the same is approximately 2/9, we get a roughly 
7\% chance to get two leptons
in a neutralino decay.  These leptons are necessarily opposite-sign (OS), but
they are flavor-uncorrelated.  This is to be contrasted with the
completely sign/flavor-uncorrelated dileptons discussed above.

Since this signal can be produced from neutralino decays, but not from
chargino decays, its size depends on their relative production rates.
For example, in the case of events that contain two binos, these OS
dileptons would account for roughly half of all dileptonic events.
Pure wino production would have a weaker signal, as 
there is only a 1/3 chance for a given side to produce $\wino^0$.
In such a case, the excess OS leptons account for about 15\% of the dileptonic events.
In both cases, the dilepton channels account for almost 20\% of all events,
with monolepton and no-lepton each accounting for $35\sim 45\%$.

We can further subdivide the dileptonic modes in the usual way, using
the relative sign/flavor of the two leptons.  
They can be either opposite-sign opposite-flavor (OSOF),
opposite-sign same-flavor (OSSF), same-sign opposite flavor (SSOF),
or same-sign same flavor (SSSF).  For the signals discussed above, 
we get flavor-universal ratios of 1:1 for OSOF:OSSF and SSOF:SSSF.  
But given the presence of our OS lepton production from neutralino decay followed
by slepton decay, there will be an excess of OSOF and OSSF compared
to SSOF and SSSF.  The numbers above suggest that this excess will be 
${\mathcal O}(0.1\sim 1)$.

In addition, there will be very distinctive
trilepton modes, where one side of the event produces OS dileptons in neutralino
decay, and the other side produces a single lepton in either chargino or neutralino decay.  
The rates for these processes are only about 2 times smaller
than the OS dilepton excess.  We will
discuss these in more detail in subsection~\ref{trilep}, but at this point we can
already see that the three leptons will be completely flavor-uncorrelated,
and that there will be no events where all three leptons have the same sign.

There will even be a population of 4-lepton events, when both chains produce two OS
leptons.  This tends to be smaller than the trileptons by about an order of magnitude.
We will not carefully investigate these events, but they will serve as further
confirmation if they are observable.

To summarize, we expect the following pattern of multilepton events when
the sneutrino is the (N)LSP:
\bit
\item  $35\sim 45\%$ of all SUSY events will have no leptons, another $35\sim 45\%$ 
will contain one lepton, and close to 20\% will be dileptonic.
\item  On top of a general sign/flavor-uncorrelated
signal from leptons produced in independent chains, the dileptonic channel will
contain a flavor-uncorrelated excess of OS events.  The relative size of this excess
depends on the relative rates of gauginos produced, but it will account for
between 15\% and 50\% of the entire dilepton sample.
\item  Trilepton events will be present, with numbers roughly half as large 
as the correlated OS excess (OS minus SS signal).
\item  4-lepton events may also be observable, though they will have much smaller rates.
\eit
We emphasize that these are just rough estimates.  In particular, phase space and mixing of
gaugino couplings can in principle modify these numbers at the $\mathcal{O}(1)$ level.

\subsection{Opposite-Sign Dilepton Excess}\label{OSdilep}

\begin{figure}[t]
\centering
\includegraphics[width=3.3in]{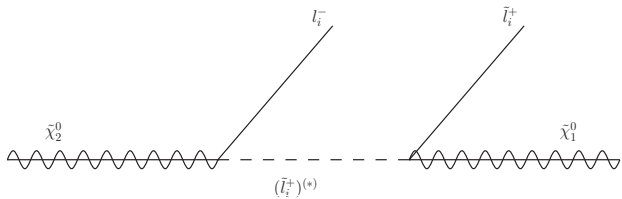}
\caption{The diagram responsible for the $\tilde{\chi}^0_2$ decay 
down to $\tilde{\chi}^0_1$ via an intermediate (possibly off-shell) 
slepton, which occurs in more conventional spectra.  The leptons are
correlated in both charge and flavor, falling into the OSSF category.}
\label{fig:3}
\end{figure}

\begin{figure}[t]
\begin{center}
\epsfxsize=0.44\textwidth\epsfbox{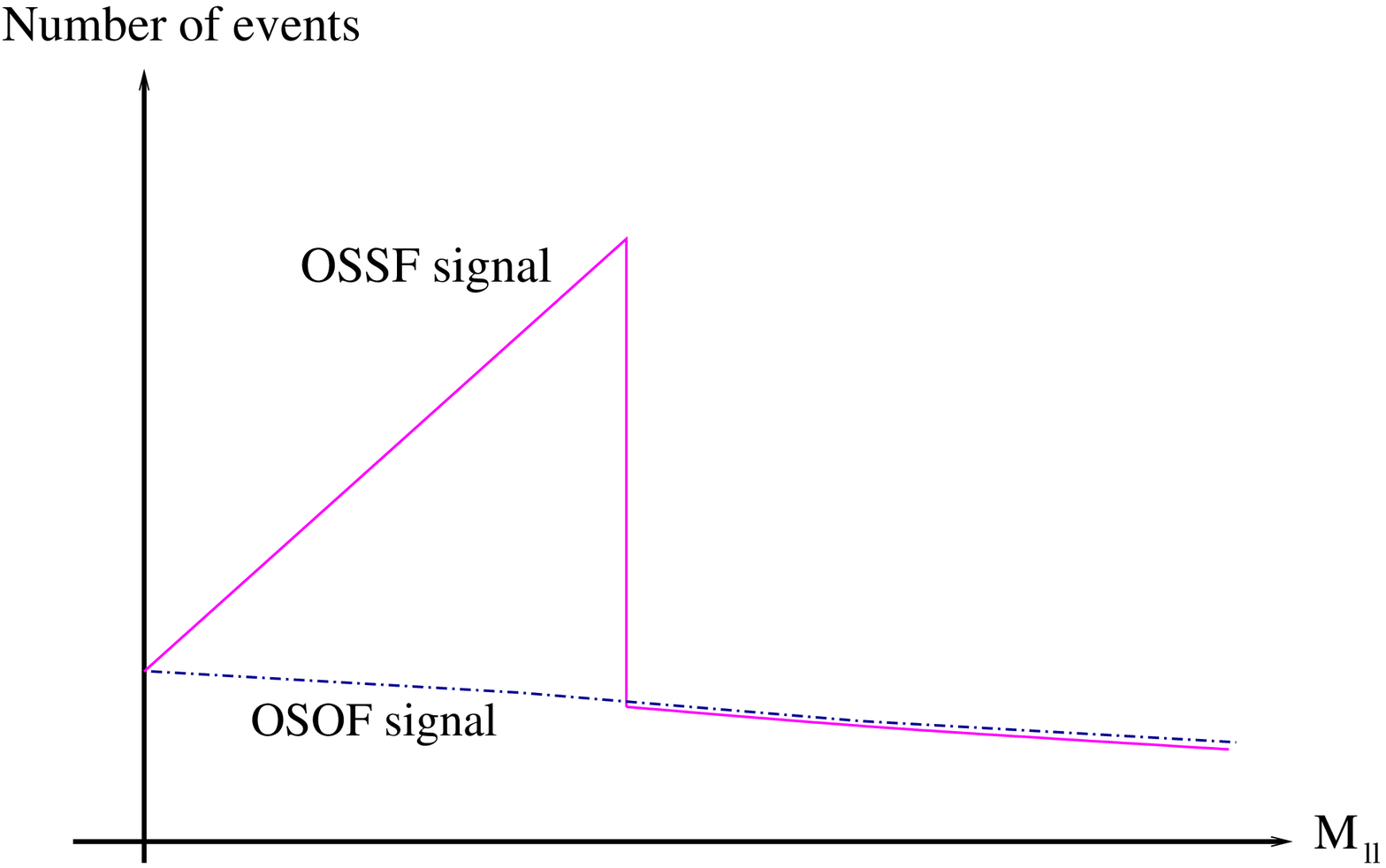}
\epsfxsize=0.44\textwidth\epsfbox{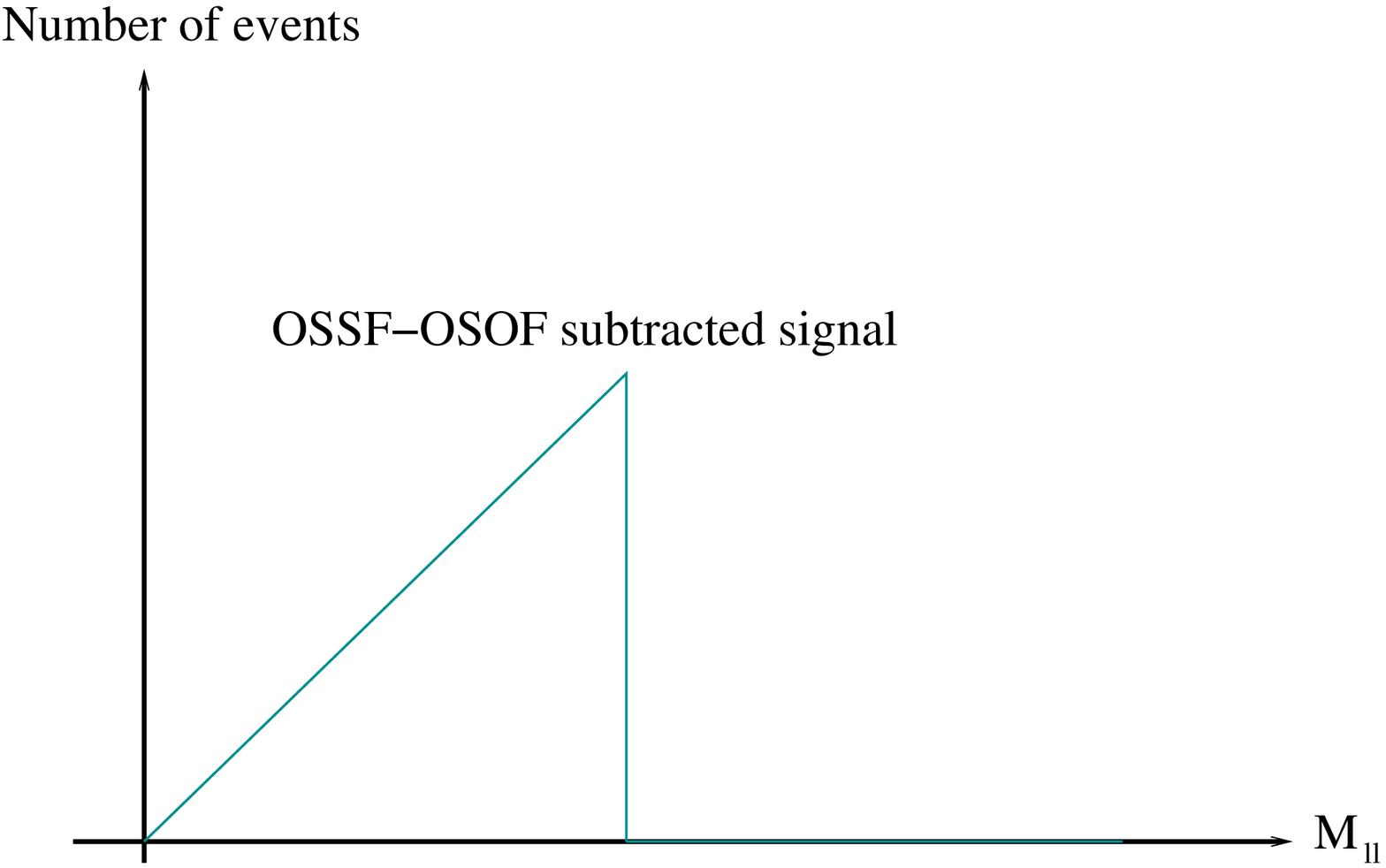}
\caption{Schematic illustration of the standard SUSY dilepton flavor subtraction.  
The left panel shows the OSSF (magenta, solid) and OSOF (blue, dot-dash) 
dilepton invariant mass distributions.  To extract the correlated contribution to
OSSF, we subtract the OSOF shape.  The right panel shows the result of the subtraction.}
\label{fig:OSSFsubtraction}
\end{center}
\end{figure}

A standard signal of many SUSY spectra is an excess of OSSF dileptons, from the 
neutralino chain depicted in Fig.~\ref{fig:3}.  This is generally accompanied by 
flavor-uncorrelated backgrounds, from both the SM and from SUSY.  If we plot the dilepton 
invariant mass distributions from both the OSOF and OSSF channels, we see purely 
uncorrelated lepton pairs for the former, and a combination of uncorrelated and 
correlated leptons for the latter (left panel Fig.~\ref{fig:OSSFsubtraction}).  
In order to isolate the 
correlated contribution, one can perform a flavor subtraction, OSSF minus OSOF 
(right panel Fig.~\ref{fig:OSSFsubtraction}).
This reveals a dilepton mass distribution which contains information about the 
mass splittings between the neutralinos and the slepton.  For example, when the slepton
is on shell, we get the characteristic ramp and edge shape.

\begin{figure}[t]
\begin{center}
\epsfxsize=0.44\textwidth\epsfbox{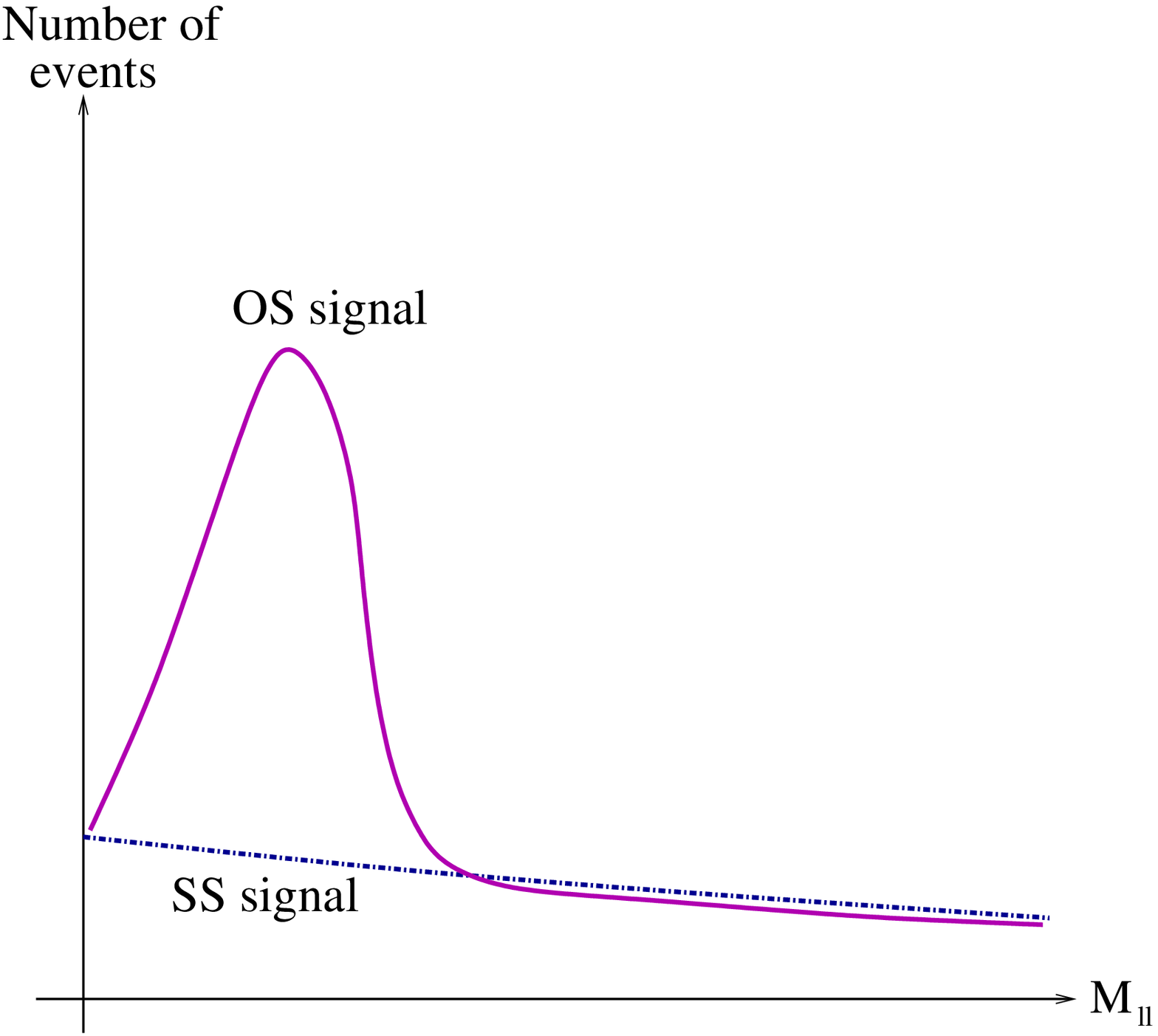}
\epsfxsize=0.44\textwidth\epsfbox{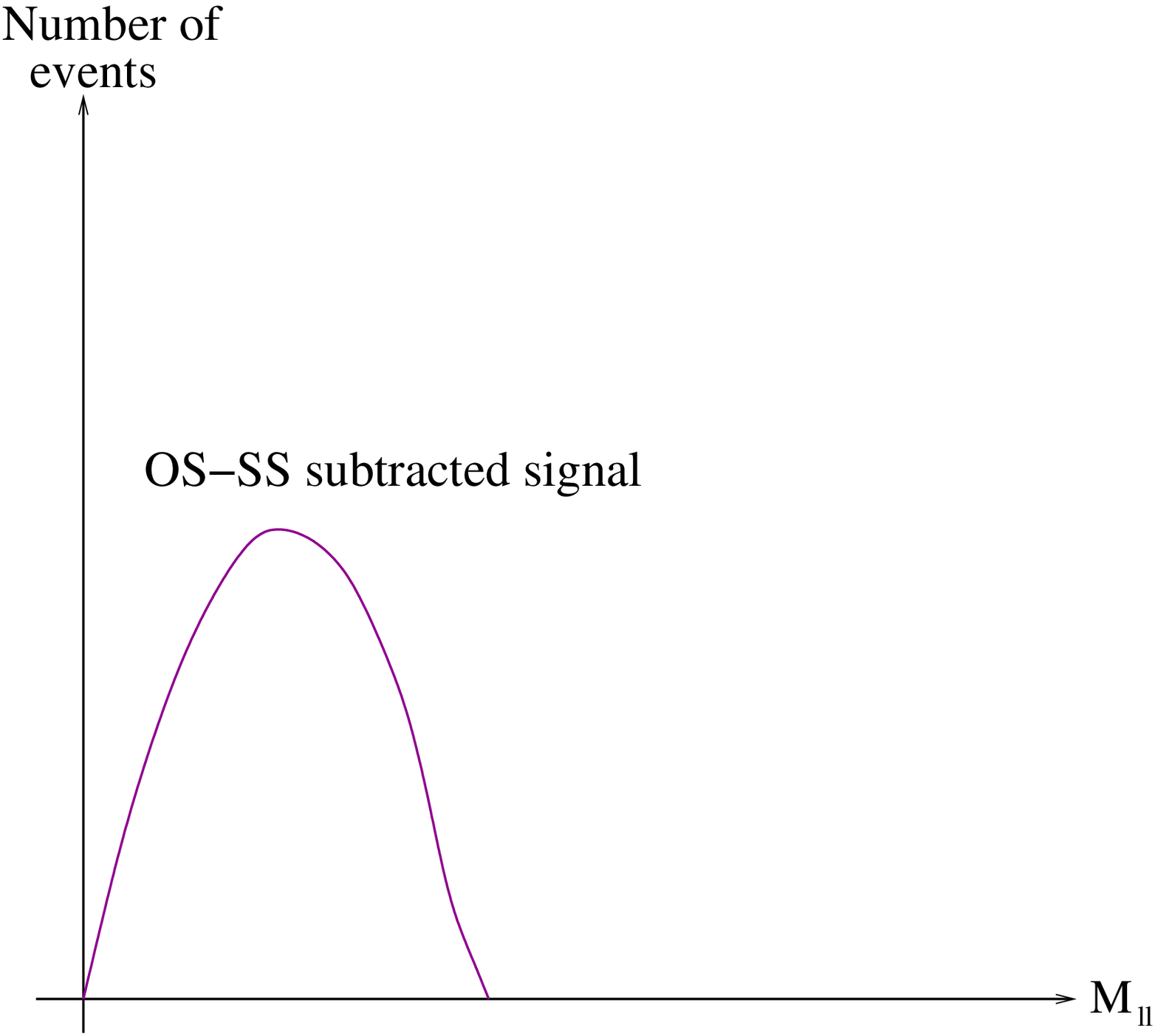}
\caption{Schematic illustration of our dilepton sign subtraction.  
The left panel shows the OS (magenta, solid) and SS (blue, dot-dash) 
dilepton invariant mass distributions.  To extract the correlated contribution to
OS, we subtract the SS shape.  The right panel shows the result of the subtraction.}
\label{fig:OSsubtraction}
\end{center}
\end{figure}

With the sneutrino NLSP spectra, we find ourselves in a very analogous situation.
As discussed above, opposite-sign uncorrelated-flavor lepton pairs are produced 
in the decays of neutralinos into charged sleptons, followed by decay down
to the sneutrino (bottom-right panel of Fig.~\ref{fig:2}).  These decays will
produce their own distinctive dilepton invariant mass distribution (discussed in
more detail below), now encoded 
in equal excesses in the OSOF and OSSF channels.  Immediately, we can
see that the above subtraction scheme will completely miss this signal.
In order to find it, one should perform a subtraction not in flavor, but in
sign.\footnote{A similar subtraction
has recently been advocated in~\cite{Kumar:2009sf}, 
in studying scenarios with mostly right-handed sneutrino
LSP in the MRSSM~\cite{Kribs:2007ac}.  However, the framework there
is genuinely flavor-violating, and incorporates different analysis tools.}  
This is illustrated in Fig.~\ref{fig:OSsubtraction}.  
In order to further test flavor universality, the subtraction can be performed in
individual opposite-flavor and same-flavor categories (OSOF minus SSOF, and OSSF 
minus SSSF).  Subsequently, we will leave 
this option implicit, and simply refer to OS and SS categories.

Two important qualifications to this procedure are in order.  First, it is possible
that the SUSY production will have sign correlations, for example if it is biased
towards squark-antisquark pairs from s-channel gluons.  In such a case, the uncorrelated
OS and SS distributions may have different normalizations.  However, they will still
have the same shape.  The OS excess could then still be revealed with a weighted subtraction,
such that the subtracted shape is left with no high-mass tail.  Second, the SM backgrounds
(mainly $t\bar t$)
are dominantly OS.  Indeed, the discovery of the SS signal will be a cleaner first 
indication that there is new physics at play.  In the standard flavor-subtraction,
these backgrounds cancel out.  In our case, they
do not.  Interpretation of the SS-subtracted OS excess as additional new physics will 
therefore require greater care.  Nonetheless, as we will show in section~\ref{signat}, the
backgrounds can be brought to a manageable level, such that the signal becomes dominant
with reasonable spectra.

\begin{figure}[t]
\centering
\includegraphics[width=3.3in]{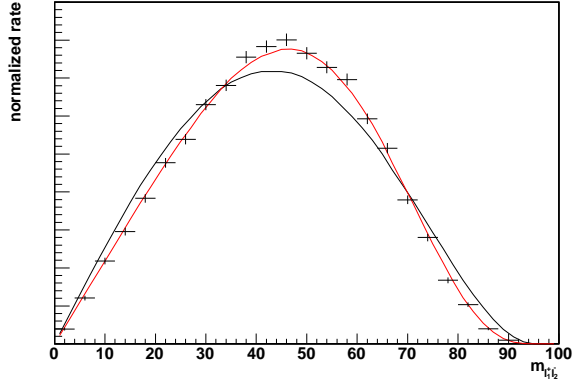}
\caption{The OS dilepton invariant mass distribution from the sequence 
$\gaugino^0 \to l_i^- \slep_i^+ \to l_i^-(l_j^+ \nu_j \snu^*_i)$, with spectrum 
$m_{\tilde \chi^0}=362\ {\rm GeV}$, 
$m_{\slep} = 232\ {\rm GeV}$, $m_{\snu} = 218\ {\rm GeV}$.
The black line shows
the distribution from flat phase space.  The red line shows the distribution
incorporating matrix elements.  The histogram shows a distribution obtained
using {\tt BRIDGE}.}
\label{fig:4}
\end{figure} 

In addition to a somewhat unconventional distribution between sign/flavor bins, the
shape of the OS excess has its own unique features.  We can first observe that
instead of a sequence of two 2-body decays or one 3-body decay, we have a 2-body
decay followed by a 3-body decay.  In fact, this possibility has already been
discussed in the context of models with mostly right-handed sneutrinos as the 
(unqualified) LSP~\cite{Thomas:2007bu}.  There it was pointed out that the dilepton
mass distribution has a shape distinct from both of the ordinary two options.
We display this shape for a sample spectrum, with and without spin effects in the
slepton decay, in Fig.~\ref{fig:4}.
(We also include a distribution obtained with 
the program {\tt BRIDGE}~\cite{Meade:2007js}, which we further utilize in the
analysis of section~\ref{signat}.)

As usual, this shape contains information on the masses (and even the spins) of the particles 
participating in the decays.  The endpoint is given by the same kind of expression 
which describes the endpoint of the $\gaugino^0_2 \to \slep \to \gaugino^0_1$ 
sequence with on-shell slepton:
\begin{equation}
m_{max} = \sqrt{\frac{(m_{\gaugino}^2-m_{\slep}^2)(m_{\slep}^2-m_{\snu}^2)}{m^2_{\slep}}} \simeq
\sqrt{ 2(m_{\gaugino}^2-m_{\slep}^2) \frac{m_{\slep}-m_{\snu}}{m_{\slep}} }~.  \label{eq:mll}
\end{equation}
We also note that in the small-splitting limit, the
entire distribution achieves a universal shape, described by a 
ninth-degree polynomial.  For details, see appendix~\ref{mlldistrib}.  The peak, which may be much 
easier to measure in practice than the endpoint, occurs near $(0.48)m_{max}$.

Several other aspects of the OS dilepton excess are worth noting:
\bit

\item  The distribution may be bimodal, corresponding to the two subchains 
$\gaugino_2^0 \rightarrow \slep \rightarrow \snu$ and 
$\gaugino_1^0 \rightarrow \slep \rightarrow \snu$.  If the two neutralinos
have similar masses, then they
may be difficult to disentangle.

\item  While the presence of the completely sign- and flavor-uncorrelated
dilepton events acts like 
a background here, its presence is also a crucial 
point of verification of the sneutrino 
NLSP interpretation.  As per the accounting in the previous subsection,
the total number of uncorrelated dilepton events will be ${\cal O}(1 \sim 10)$ 
larger than the observed OS excess.

\item  As we have already mentioned, the slepton/sneutrino mass 
difference scales inversely with the average doublet mass, and can 
be relatively small compared to the 
superpartner mass scale.  Consequently, the lepton produced in the slepton decay
will not necessarily have much available energy, and it must share this
with the neutrino and (to a much lesser extent) the sneutrino.  
This can complicate the clean identification 
of the lepton, which on average acquires an energy
of approximately $(m_{\slep}-m_{\snu})/2$ in the slepton rest frame.
(For example, for a 200 GeV doublet, the average energy is about 8 GeV.)
With strict identification requirements to reject fakes, and
tight isolation cuts to reject heavy flavor decays, ${\cal O}(1)$
of these leptons may be unusable.  We will demonstrate that a significant 
fraction can still be used in realistic cases in section~\ref{signat}.

\eit

\subsection{Trileptons} \label{trilep}

Whenever one side of the event produces an OS dilepton pair, the other side has 
$\mathcal{O}(1/3)$ chance of also producing a lepton via gaugino decay.  This leads
to a roughly 2:1 ratio between rates for correlated OS dilepton events and very 
distinctive trilepton events, which have smaller SM backgrounds.  If the OS excess 
is generated with high enough statistics
to be clearly observable, the trilepton events will be observable as well.  Indeed, the presence
of these trilepton events may very well be established before the OS minus SS subtraction 
is possible.

Beyond simple counting, it should also be possible to extract the correlated dilepton
mass distribution (Fig.~\ref{fig:4}) from these events, leading to a powerful confirmation
of the overall picture.  However, there is an immediate combinatoric problem to be overcome.
The only sign/flavor structure between these three leptons is that they cannot all
have the same sign.  In general, in every event we will be able to find \emph{two} possible OS 
dilepton
pairings, only one of which is the correct one.  This is to be contrasted with the 
analogous production of trileptons in more standard SUSY spectra, 
where we are always guaranteed an OSSF pair.  There we
can focus on events where this pair is generated with a lepton of the opposite flavor.

For our case, we propose a simple strategy which breaks the degeneracy and directly
extracts the correct distribution to good approximation.  The lepton emitted in the slepton
decay will tend to be the softest lepton in the event.  \emph{Assume} that this is indeed
the case.  Focus on the subsample of events where only one of the hardest two leptons
can form an OS pair with this soft lepton.  (Equivalently, the two hardest leptons should
be OS with respect to each other.)  
When we form the invariant mass distribution from this (half-sized) subsample, 
we should have a much higher probability of
having guessed correctly, and the mass distribution should reflect this.  We
find that this technique works well in practice, and demonstrate its application
in section~\ref{signat}.

The trilepton signal could also be useful for discovery via electroweak production
of gaugino pairs.  As usual, the high lepton multiplicity can make the events
distinctive enough that SM backgrounds can be controlled, even though the accompanying
jet activity and missing energy are not necessarily very large.  We will not explicitly 
investigate this discovery channel here, but note that the observations of this
section clearly still apply.  In particular, the signal can immediately be discriminated
from typical trilepton SUSY production by the complete lack of flavor correlations.

It is possible that the Tevatron is already capable of placing interesting limits
on sneutrino NLSP spectra in EW production.  We save investigation of this for
future work.

\subsection{Jet-Lepton Invariant Mass} \label{mjl}

As in more traditional searches, we can also attempt to construct kinematic
observables from the jets produced in the cascades.  In particular, the
sequence $\squark \to q\gaugino \to ql(\slep/\snu)$ results in a jet and a 
lepton, from which we can construct an invariant mass distribution.
Since we assume that the gaugino is on-shell, this distribution has the 
same ramp-and-edge shape that characterizes the dilepton mass distribution
in $\gaugino^0_2 \to l^+\slep^- \to l^+l^-\gaugino^0_1$ (right panel
Fig.~\ref{fig:OSSFsubtraction}).\footnote{Because the intermediate particle in
the decay chain is now spin-1/2, there will be spin effects modifying the distributions
from flat phase space.
However, these average out when we sum over quark and lepton charges~\cite{Miller:2005zp}.}  
Extraction of such a shape would help further confirm the origin of the
leptons.  It would also provide information on the squark/gaugino mass splittings
since the edge is predicted to occur as in equation~\eqref{eq:mll} with appropriate
replacements.

Of course, reality is not so straightforward.  Even assuming high quality
jet measurements, we are always left with combinatoric ambiguities since
each side of the event will produce jets.  Still, even incorporating
multiple pairing possibilities into the construction of the mass distribution (or
simply guessing at random), nontrivial edge features will remain.
With high enough statistics, we may be able to cleanly identify
these edges, recovering some kinematic information and achieving
greater confidence that we are seeing leptons produced in two-body gaugino 
decays.\footnote{It was
also suggested in~\cite{Thomas:2007bu} that a kind of subtraction, analogous to what
is done for dilepton distributions, might be possible in order to remove
the shape contribution from uncorrelated jet-lepton pairings.}

In events with gluino decays, these edges can be washed out, or
disappear entirely.  A heaver gluino decaying via an on-shell squark
injects an extra hard jet into the event, further complicating the combinatorics.
A lighter gluino decaying via off-shell squark effectively destroys the
ramp-and-edge structure.  In cases where gluino production dominates, there
may therefore be no edge-like features in the jet-lepton invariant mass distribution.

It should also be noted that we may accidentally use a lepton produced in slepton
decay, versus gaugino decay, in the jet-lepton mass.  However, this 
is not a major issue.  Monoleptonic events are already dominated by leptons produced
in gaugino decay (by a factor of $3 \sim 6$), and can be further
purified by demanding harder leptons.  In multilepton events, we usually
make a correct choice by taking the hardest lepton.

\subsection{How Unique are These Signals?}\label{fakers}

Will these signals be enough to confidently infer that the 
decay chains are ending in sleptons/sneutrinos?  In order to address this question,
let us consider how leptons are produced in alternative scenarios within the MSSM.

SUSY cascades primarily generate leptons in three ways:  decays of gauginos into
sleptons/sneutrinos, decays of sleptons/sneutrinos into gauginos, and decays
of electroweak gauge bosons emitted in transitions between charginos/neutralinos.
Any of these particles could also be off-shell.  In sneutrino NLSP spectra, a fourth
option opens up, from the decays $\slep \to \snu l \bar \nu$ discussed in 
subsection~\ref{decays}.  This is a crucial component of our signal.  
But when sneutrino is not at the bottom of the spectrum, 
these three-body decays are bypassed in favor of the two-body decays 
$\slep^+ \to \tilde \chi^0 l^+$ or $\slep^+ \to \tilde \chi^+ \nu$.

The most commonly considered way to produce multiple leptons in the
same chain is $\tilde{\chi}^0_2 \to \tilde{\chi}^0_1$  
via an intermediate slepton,
as shown in Fig.~\ref{fig:3}.  This leads to a very distinctive excess in the
OSSF dilepton channel, typically constituting an ${\cal O}(1)$ fraction of
all dilepton production.  Such an excess will not be present with our 
sneutrino NLSP unless we also introduce the RH slepton into the cascades.  
(But this has its own set of novelties, which we will discuss in~\cite{active}.)  

This OSSF excess is a very generic signal whenever charged sleptons participate 
in the decay 
chains.\footnote{An obvious way around this is to have large 
flavor violation in the slepton 
sector.  This is clearly dangerous from the perspective of 
$\mu \rightarrow e\gamma$, 
so we assume flavor universality in our own analysis.  
But for some interesting ways to have large SUSY 
flavor violation while avoiding disagreement with low energy 
precision experiments, 
see~\cite{Feng:2007ke,Nomura:2007ap} and~\cite{Kribs:2007ac}.  In order to
fake our flavor-uncorrelated signal, this non-universality would need to be 
nearly maximal, but this possibility is excluded 
(see~\cite{Nir:2007xn} for review).} 
But it can be hidden if there is an accidental mass degeneracy with one
of the charginos/neutralinos, such that one of the emitted leptons becomes very difficult
to find.  Assuming this occurs, then we are left with 
uncorrelated production, where the leptons in dileptonic events always
come from opposite chains.  If the sleptons are rarely bypassed, 
this could lead to a similar dilepton:monolepton
ratio.

Another way to get single leptons in each chain is through emission of $W^{(*)}$,
from a $\tilde{\chi}^+ \to \tilde{\chi}^0$ transition, or vice-versa.  The branching
fraction into electron and muon is 22\%, which is not so different from our 33\%
branching fraction for gaugino decays into sleptons.  The counting may therefore
be different only at the $\mathcal{O}(1)$ level, assuming the $W^{(*)}$ is produced
in most decay chains.

While each of these two possibilities -- slepton transitions with a near-degeneracy
and $W^{(*)}$ emission -- do not lead to correlated dileptons within a single
chain, there may nonetheless be sign correlations across an event originating at production.
(Again, domination by squark-antisquark production is a simple example.)  The naive
sign-subtraction procedure in the dilepton invariant mass distribution will leave 
over an OS excess, but, unlike the sneutrino NLSP case, this will have
exactly the same shape as the SS distribution (up to SM backgrounds).  In addition, if
we only get one lepton produced in each of the two decay chains in an event, we
will never get a trilepton excess.  In order
to get either a distinctive OS excess or trileptons, we generally need the possibility 
of two leptons being produced in the same chain.

To get multiple lepton emissions, we can string together these types of transitions.
Indeed, a $W^*$ emission in association with slepton production or decay could closely 
mimic the pattern of our signal, particularly if there is a modest degeneracy 
analogous to the slepton/sneutrino mass splitting.

While these are indeed logical possibilities, they are clearly highly
non-generic.  Besides the requirement of strong accidental mass degeneracies to
hide the OSSF signals whenever sleptons contribute, at least 
three distinct mass levels of charginos/neutralinos must be participating in almost
every decay chain.  This might become possible when there is large mixing, or 
if the mostly-Higgsinos are much lighter than the mostly-gauginos.  We will not 
investigate these alternatives in detail, since, in any case, there are too many variations 
to explore exhaustively.  
But these arguments do
suggest that the pattern of leptons in spectra with a sneutrino NLSP is quite
distinctive.

\section{LHC Simulations} 
\label{signat}

In this section we will use the tools developed in
section~\ref{cascades} in order to analyze two example spectra.
We perform simulations including
showering/hadronization, event reconstruction at particle level, and SM backgrounds.
We do not attempt to incorporate a detector simulation or energy resolution
effects.  The purpose here 
is to demonstrate the plausibility of a real search at the LHC, given
the presence of backgrounds and a jetty environment where leptons might
be lost, or simply fall below the energy threshold of good quality
reconstruction.  This last point is particularly relevant for the leptons produced
in slepton decay down to the sneutrino, which can be somewhat soft due to the small
mass splitting.

In the following, we will assume that the LHC will ultimately reach its design
energy of 14 TeV.  We do not expect that a lower final operating energy will qualitatively 
change our conclusions.  For example, at a 10 TeV LHC, the signal cross sections are reduced by
a factor of about $3\sim4$.

\subsection{Sample Spectra}\label{specsim}

\begin{table}
 \centering
\begin{tabular}{|c|c|c|c|c|c|c|c|c|c|c|c|c|c|}
 \hline
  & Input scale & $M_{\bino}$ & $M_{\wino}$ & $M_{\gluino}$ & $m_{\sel}$ & $m_{\slep}$ & $m_{\tilde Q}$ & $m_{\tilde \bar u}$ &
$m_{\tilde \bar d}$ & $m_{H_u}^2$ & $m_{H_d}^2 $ & $\tan \beta$ & sign $\mu$ \\  \hline
 $\tilde q$GGM & $10^5$ & 350 & 370 & 1200 & 350 & 200 & 499 & 539 & 500 & -60000 & 60000 & 10 & + \\
 $\tilde g$GGM & $10^5$ & 350 & 370 & 600 & 352 & 200 & 1500 & 1514 & 1500 & -60000 & 60000 & 10 & + \\ \hline  
\end{tabular}
\caption{Input parameters for the sample spectra in units of GeV (${\rm GeV}^2$
for the masses squared).}
\label{tab:input}
\end{table}

There are many possible spectra with sneutrino NLSP that we might consider, but
most of them lead to qualitatively similar phenomenology if we restrict ourselves
to the assumptions outlined in section~\ref{cascades}.  In particular, the mass ordering
between the wino and bino is not very important, as long as they are lighter than the
squarks and gluinos.  Other than very detailed
variations involving large gaugino-Higgsino mixings, or introduction of approximate
mass degeneracies, the main flexibility available to us is the mass of the LH
slepton doublet and the production of the gauginos in squark/gluino decays.  The 
former controls the mass splitting between the slepton and sneutrino, and hence
the energy of the leptons produced in slepton decays.  The relative rates
of bino versus wino production will control the size of our OS dilepton excess
and trilepton signal.  Weaker signals occur if winos dominate, since fewer neutralinos
will be produced.  
In addition, the proportion of gauginos produced directly in squark 
decays versus in gluino decays will determine the visibility of edge structures
in the jet-lepton invariant mass distribution.

We concentrate on two SUSY spectra with sneutrino NLSP within the framework of
GGM with arbitrary Higgs sector.  The spectra have similar slepton and gaugino
masses, but the ordering of squarks versus gluino is different.  In the first
spectrum, labeled ``$\sq$GGM,'' the physical squark masses are just under 1 TeV, while
the physical gluino is about $1.4$ TeV.  SQCD production is dominated by squarks,
and decays into winos and binos are roughly democratic.  In the second spectrum,
``$\gluino$GGM,'' the situation is approximately reversed, with an 800 GeV gluino
and 1.5 TeV squarks.  Production in this spectrum is dominated by gluinos,
which decay through off-shell squarks.  These decays are highly biased towards
winos, due to the larger $SU(2)_L$ couplings.\footnote{SQCD does not discriminate 
between squark chiralities, but the diagrams with off-shell squarks are weighted 
by the couplings into the final electroweak gauginos.  This is to be contrasted with 
gluino decays into on-shell squarks, where both chiralities are produced equally,
and the couplings to winos and binos simply determine the squark decay rates.  This
latter observation also applies to the $\sq$GGM spectrum for events with gluinos.}  
A full list of 
input parameters for both spectra is given in table~\ref{tab:input}.

\begin{table}[t]
 \centering
\begin{tabular}{|c|c|c|}
\hline
   & $\tilde q$GGM & $\tilde g $GGM \\ \hline 
$\gluino$                    &  1403  &   823  \\
$\tilde{u}_L/\tilde{d}_L$    &   934  &  1555  \\ 
$\tilde{u}_R$                &   956  &  1568  \\
$\tilde{d}_R$                &   934  &  1555  \\
$\tilde{t}_1$                &   877  &  1454  \\
$\tilde{t}_2$                &   954  &  1512  \\
$\tilde{b}_1$                &   908  &  1500  \\
$\tilde{b}_2$                &   934  &  1553  \\ \hline
$\gaugino^0_1$               &   301  &   313  \\
$\gaugino^0_2$               &   338  &   362  \\
$\gaugino^0_3$               &   451  &   754  \\
$\gaugino^0_4$               &   484  &   764  \\
$\gaugino^+_1$               &   333  &   362  \\
$\gaugino^+_2$               &   483  &   769  \\ \hline
$\tilde{e}_R$                &   361  &   362  \\
$\tilde{l}_L^+$              &   235  &   232  \\
$\tilde{\nu}$                &   221  &   218  \\
$\tilde{\tau}_1$             &   233  &   227  \\
$\tilde{\tau}_2$             &   361  &   364  \\ \hline
$h$                          &   112  &   115  \\
$H^0/H^+/A^0$                &   520  &   800  \\ \hline
\end{tabular}
\caption{Physical masses (in units of GeV) in the example spectra.}
\label{tab:output}
\end{table}

We used {\tt SOFTSUSY v3.0.7}~\cite{Allanach:2001kg} to extract the physical mass
spectrum from our messenger-scale parameters.  These are presented in 
table~\ref{tab:output}.\footnote{Note, that while the spectrum $\tilde g$GGM is perfectly
viable, the spectrum $\tilde q$GGM has 112 GeV Higgs, which is excluded in the Higgs-sector
decoupling limit. We use this spectrum for illustration purposes only.  Uplifting
the Higgs to the experimentally allowed range is not difficult~\cite{Dine:2007xi},
and practically will not change our conclusions (but note that other particles in the 
Higgs sector might be modified).}

\subsection{Generation and Reconstruction of Events} \label{creev}

We utilized several programs in order to generate and reconstruct our signal events.  These include
\bit

\item {\tt MadGraph/MadEvent v4.3.0}~\cite{Alwall:2007st} for generation of generic 
$2 \to 2$ SUSY pair production in 14 TeV $pp$ collisions.  (100k events for each spectrum)

\item {\tt BRIDGE v2.15}~\cite{Meade:2007js} for calculating branching ratios of 
SUSY particles and simulating decay chains.

\item {\tt PYTHIA v6.4.14}~\cite{pythiamanual} interface for {\tt MadGraph} to shower and hadronize the events.

\item {\tt FastJet v2.3.4}~\cite{Cacciari:2005hq} for jet reconstruction of the particle-level events.

\eit

After hadronization, event reconstruction proceeds as follows.  
We separate out leptons (electrons and muons) with $p_T$
above 5 GeV and $|\eta| < 2.5$,
and check them for isolation.  We scalar-sum the $p_T$ of the lepton with the $p_T$s of 
all other non-leptonic (and non-invisible)
particles within an $\eta$-$\phi$ cone of size $0.4$.  If the lepton constitutes 90\% or more
of the total $p_T$, then we consider it ``tight.''  Failing this, if the $p_T$ of the other particles
tallies to less than 10 GeV, we consider it ``loose.''  (This second class of leptons will be
be used to keep more signal in trileptonic events.)  
We set aside leptons which fail both of these criteria for clustering into jets.

\begin{figure}[t]
\centering
\epsfxsize=0.44\textwidth\epsfbox{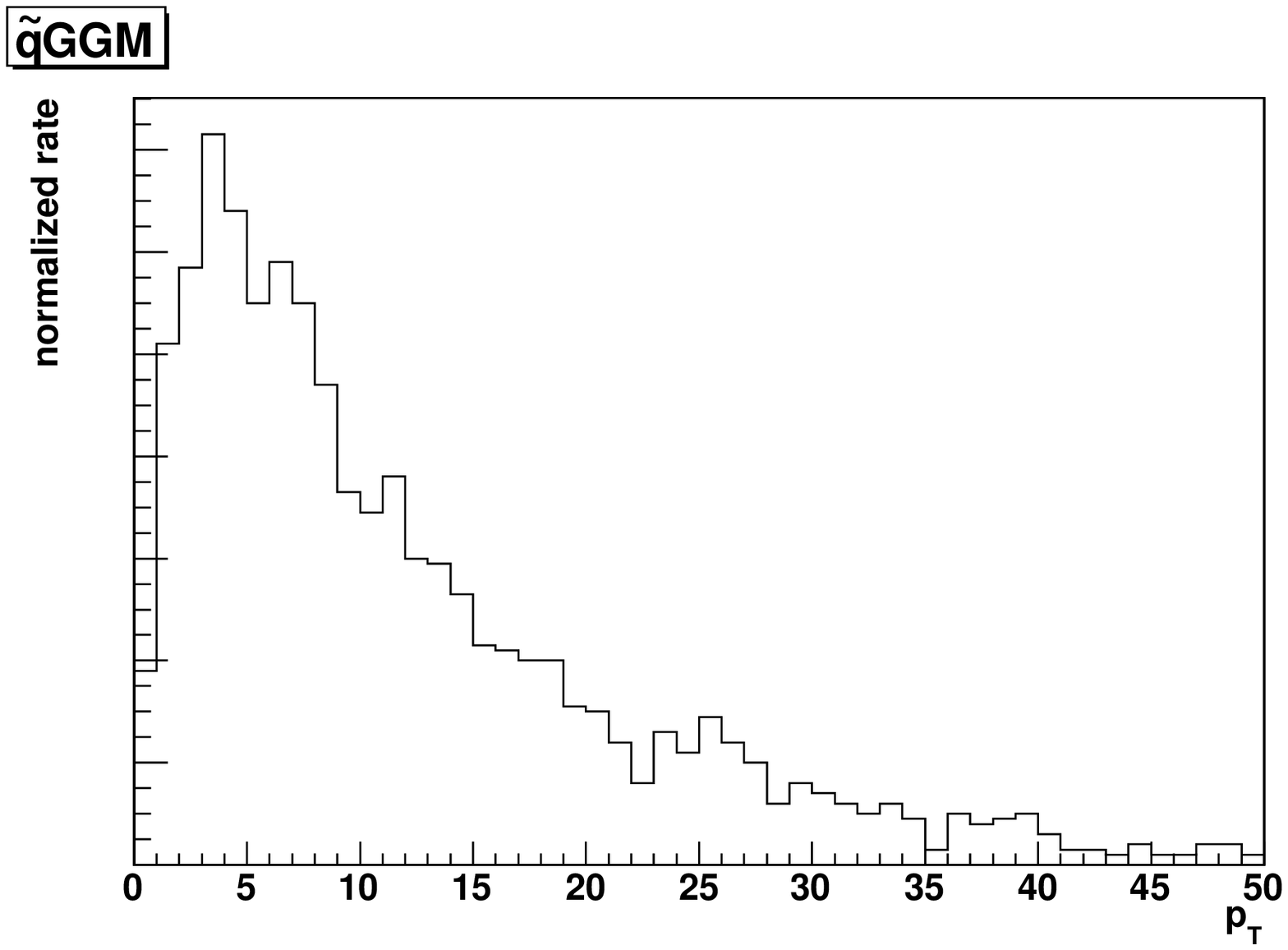}
\epsfxsize=0.44\textwidth\epsfbox{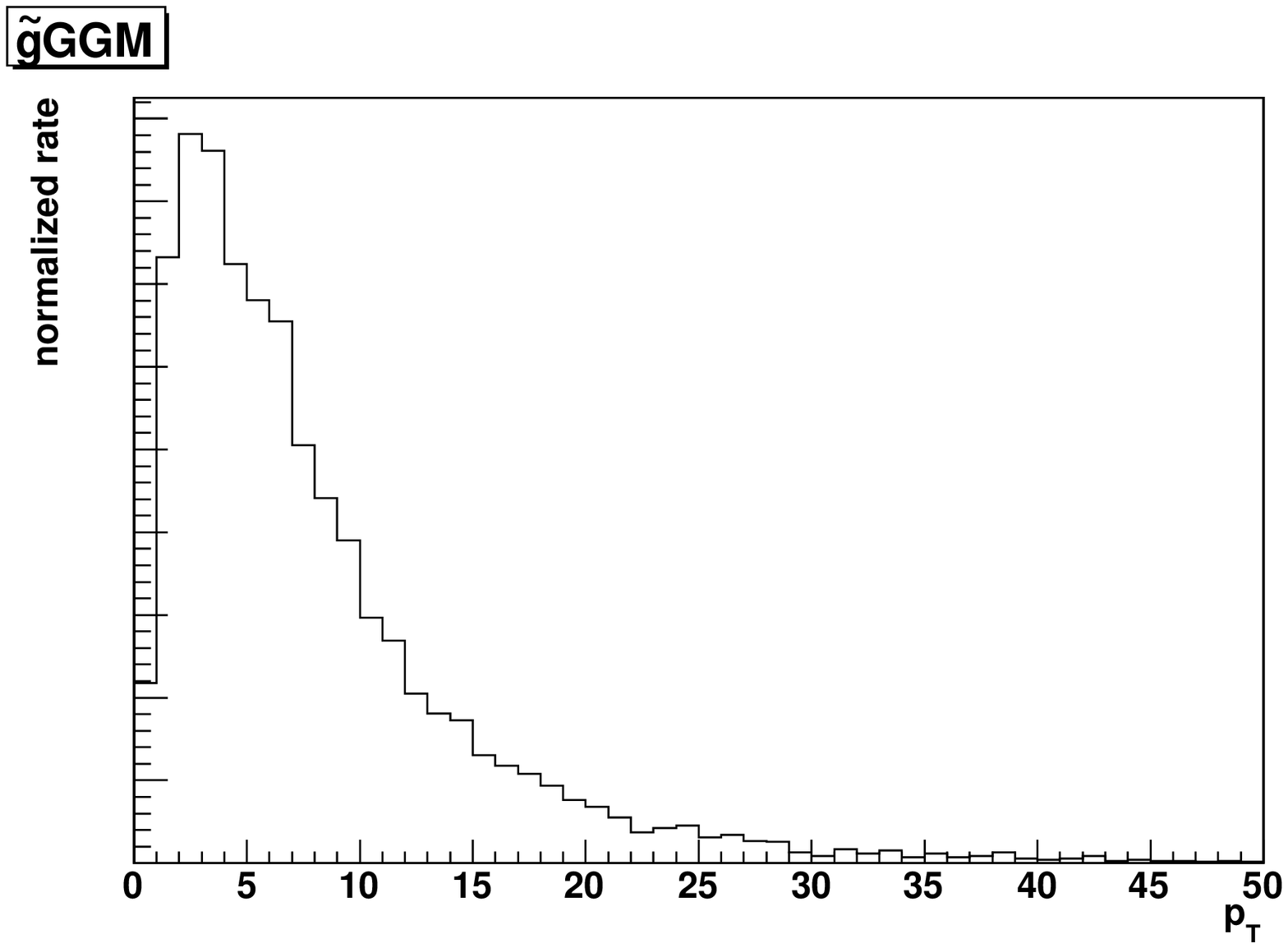}
\caption{The parton-level $p_T$ spectrum of leptons produced in three-body slepton decay, for
the $\squark$GGM spectrum (left panel) and $\gluino$GGM spectrum (right panel).  In the slepton rest 
frame, the lepton energy ranges from 0 to 14 GeV, with an average of about 7 GeV.}
\label{fig:ptlsoft}
\end{figure}

In the spectra discussed above, nearly 90\% of leptons produced in gaugino decay will be
identified as tight, and most of the rest as loose.  The leptons produced in the three-body
slepton decays will be softer, and therefore more difficult to identify and isolate.
In particular, we lose up to 30\% of them due to our 5 GeV $p_T$ threshold.
(See Fig.~\ref{fig:ptlsoft} for the complete $p_T$ distributions.  Note the large
tails due to the boost of the sleptons.)
The efficiency for detecting the leptons that pass the threshold depends on the amount
of jet activity in the event, varying from almost unity for squark pair production 
down to about 50\% for gluino pair production.  $\mathcal{O}(50\%)$ of these
leptons are identified in the loose category.

After identifying the set of isolated leptons, we proceed to cluster all of the remaining
non-invisible particles
in the event into jets using the Cambridge/Aachen algorithm with $R = 0.4$.  We keep jets with
$p_T > 20$ GeV and $|\eta| < 2.5$.

Since we will focus on heavy SQCD production events, we apply cuts on jet activity and missing 
energy.  We demand at least
two jets above 300 GeV of $p_T$, and \met\ of at least 200 GeV.\footnote{These cuts have not been optimized.
A more complete analysis may achieve better signal vs background discrimination.  For example, we might 
demand larger jet multiplicities but with a looser $p_T$ cut.  
Such a cut could be more inclusive of gluinos decaying through off-shell squarks.}
  We do not consider events with
zero leptons, though these would of course be interesting to investigate in a more complete analysis.
We consider events with one lepton if it is tight, and if the transverse
mass of the lepton and missing energy vector is above 100 GeV, to veto $W$ backgrounds.  
An event with more
than one lepton must have at least two tight leptons, or else we discard it.
In particular,
trilepton events may have a single loose lepton.  This keeps $\mathcal{O}(50\%)$ more of the
signal in that channel, while the backgrounds remain small.  We also neglect events with
any OSSF dilepton pairs between 80 GeV and 100 GeV, in order to avoid incorporating $Z$s
(either from backgrounds or produced in SUSY cascades) into our analysis.

We present leading-order cross sections for the $\squark$GGM and $\gluino$GGM spectra, 
before and after reconstruction and cuts, in appendix~\ref{tables}.

\subsection{Backgrounds} \label{backgrounds}

We simulate the leading backgrounds from SM processes that generate hard leptons and neutrinos through
electroweak boson decays.  These include
\bit

\item $t \bar t$+jets.  Matched using $k_T$-MLM at 20 GeV, up to two additional jets.  (6M events, 
$1.9$M after matching veto)

\item $WWjj$ (opposite- and same-sign).   In {\tt MadGraph}, we use cuts $\Delta R_{jj} > 0.4$, $p_{Tj} > 150$ GeV.  (70k events)

\item $l\nu jj$ via on-shell $W$ (including $\tau$s).  $\Delta R_{jj} > 0.4$, 
$p_{Tj} > 150$ GeV, \met\ $> 150$ GeV.  (500k events)

\item $\tau^+\tau^-jj$.  $\Delta R_{jj} > 0.4$, $p_{Tj} > 150$ GeV.  (500k events)

\eit
We decay the $t \bar t$ and $WWjj$ samples with {\tt BRIDGE} both in semileptonic and in
dileptonic channels, including taus.  The all-hadronic mode was found to give negligible
contribution given our cuts above.

There will also be contributions from 
$(l^+l^-)$+jets (in addition to the tau mode above), 
$(l^+l^-)W$+jets, and $(l^+l^-)(l^+l^-)$+jets.
None of these backgrounds are expected to be significant, and they are indeed found
to be subdominant in other SUSY analyses (see, e.g.,~\cite{Aad:2009wy}).
We have explicitly checked $(l^+l^-)W$+jets, which should be
the dominant electroweak background in the trilepton channel, in particular
$(\tau^+\tau^-)W$+jets in light of our transverse mass cut.  
We have verified that it is small,
with a total cross section in the trileptonic channel, after cuts, of 0.04 fb.  We do not
explicitly include it in the analysis. 

The monoleptonic channel may be polluted by generic QCD events with heavy flavor decays.
However, our cuts on tightness of the lepton and on \met\ will significantly reduce these.
We do not investigate this background here, but note that generic QCD has
been shown to be negligible in other monolepton SUSY analyses (see again~\cite{Aad:2009wy}).
These usually use a higher $p_T$ cut on the lepton, but we do not find that this has
a significant effect on our results below.

There is an additional subtlety with the monoleptonic channel.  Without additional activity
from heavy flavor,
events with a single leptonic $W$ decay will almost never pass our 100 GeV cut on transverse mass
(section~\ref{creev}).  In a more realistic simulation, most of the passing events would probably
come from the resolution tail of the detector.  However, we do not incorporate any resolution
smearing in our nominal analysis.  To get some sense of how large of an effect this might be,
we smeared the transverse mass by 20\% and checked how many events pass the cut.  The signal
is only marginally affected, but the surviving semileptonic $t \bar t$+jets, $Wjj$, and semileptonic $WWjj$
backgrounds increase by factors of roughly 3, 5, and 10, respectively.  These are still quite
small with respect to the signals.  (They can also be significantly attenuated with modest
additional cost to the signal by simply using a higher transverse mass cut.)

Leading-order cross sections for the backgrounds, before and after reconstruction and cuts, can
be found in appendix~\ref{tables}.

\subsection{Analysis of Events}\label{fullan}

Here we present the results of the analyses of section~\ref{cascades}, as applied
to our two sample spectra, $\squark$GGM and $\gluino$GGM, incorporating backgrounds.  
Events are weighted to correspond to 100 fb$^{-1}$ of integrated luminosity.  While pure
counting in multiple channels will already indicate the presence of SUSY (and even suggest
the presence of a sneutrino NLSP) long before
this amount of luminosity is acquired, we choose to present this longer-term goal so that statistics
are good enough to clearly see the shapes in all of our proposed kinematic distributions.

Indeed, if lepton fakes and missing energy are well-understood early on, a clear excess of
events in the monoleptonic channel may already be visible with as little as a few 100 pb$^{-1}$
of data, even at the planned initial operating energy of 10 TeV.  Subsequent progress will
depend somewhat on how soon 14 TeV becomes available, with $3\sim4$ times higher signal
cross sections.\footnote{This factor applies both before and after our cuts 
on the signal.  We have
not explicitly investigated the backgrounds at 10 TeV, but do not expect them to become
significantly more important relative to the signal.}  In any case, relatively background-free 
samples of same-sign dilepton and trilepton events would
start to become available with a few fb$^{-1}$.  Opposite-sign dileptons should also emerge with
statistical significance over backgrounds around this point.  The presence of a near-flavor-universal
excess of OS over SS could probably be inferred by a few 10's of fb$^{-1}$, with shape information
becoming progressively better above that.
Of course, spectra with 
lighter colored superpartners would have higher rates, and might require less running to achieve 
the same level of statistics.

\begin{figure}[t]
\begin{center}
\epsfxsize=0.49\textwidth\epsfbox{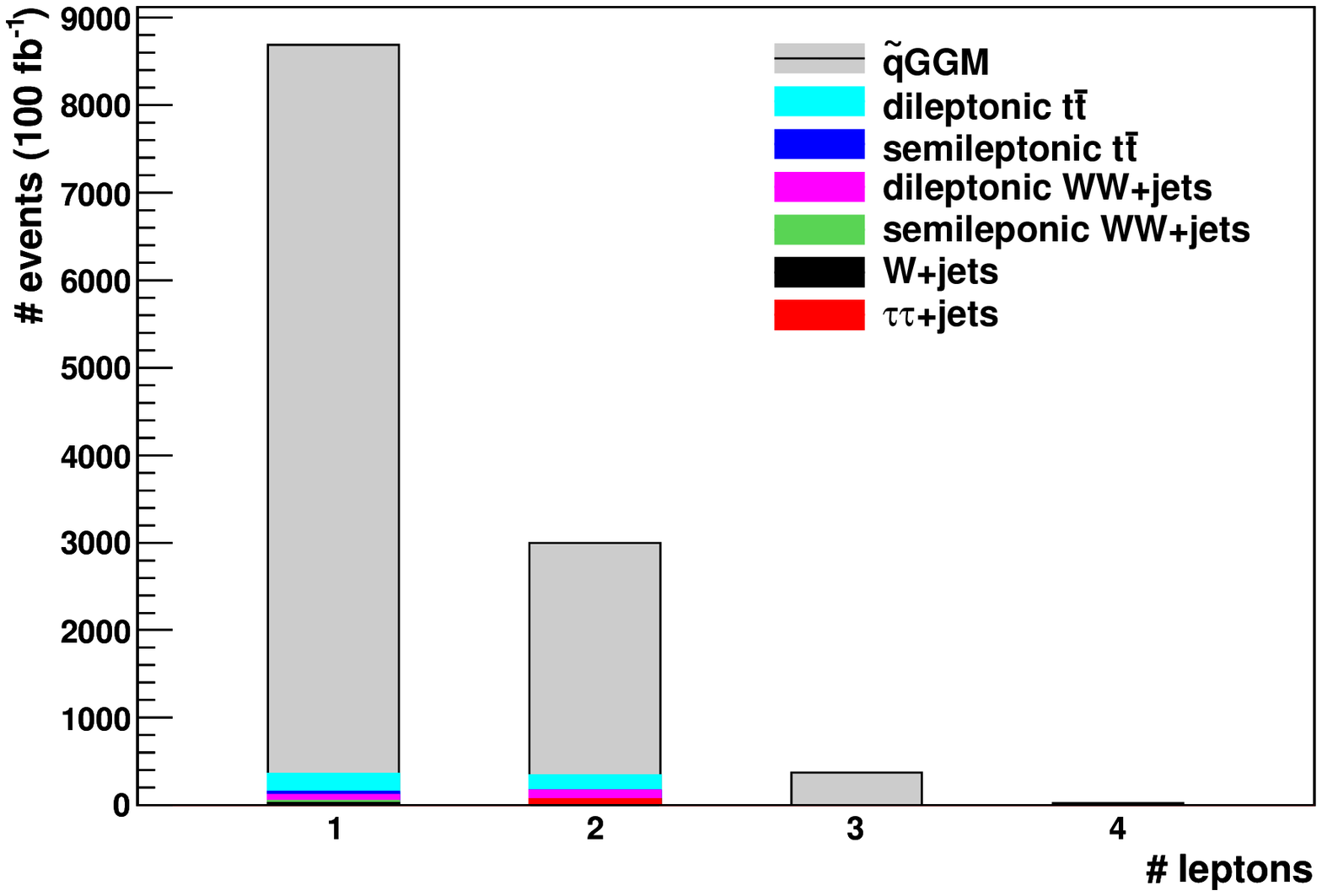}
\epsfxsize=0.49\textwidth\epsfbox{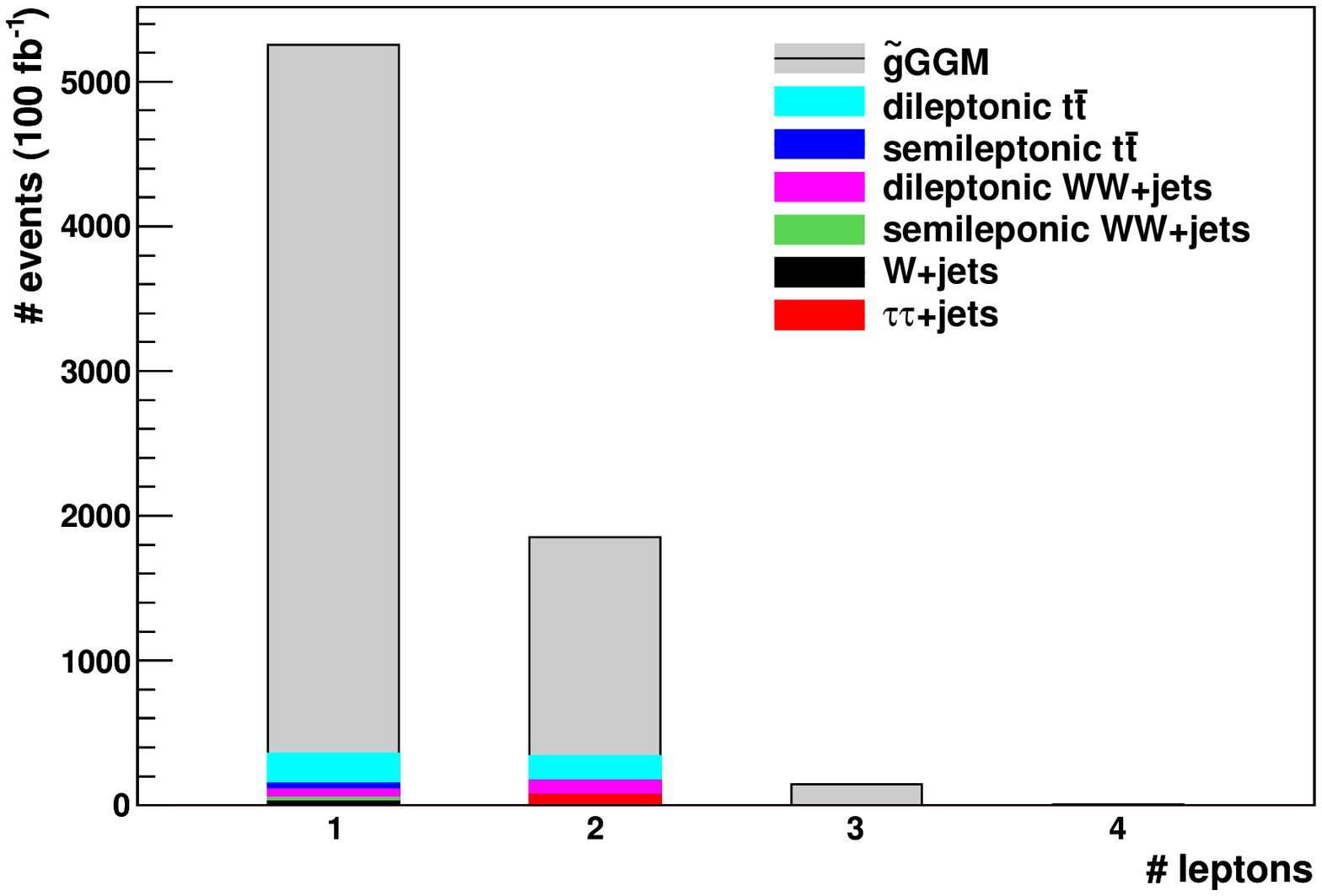}
\caption{Counts for number of observed leptons for the $\squark$GGM spectrum (left panel)
and $\gluino$GGM spectrum (right panel).  (The histograms are stacked.)}
\label{fig:nl}
\end{center}
\end{figure}

\begin{figure}[t]
\begin{center}
\epsfxsize=0.49\textwidth\epsfbox{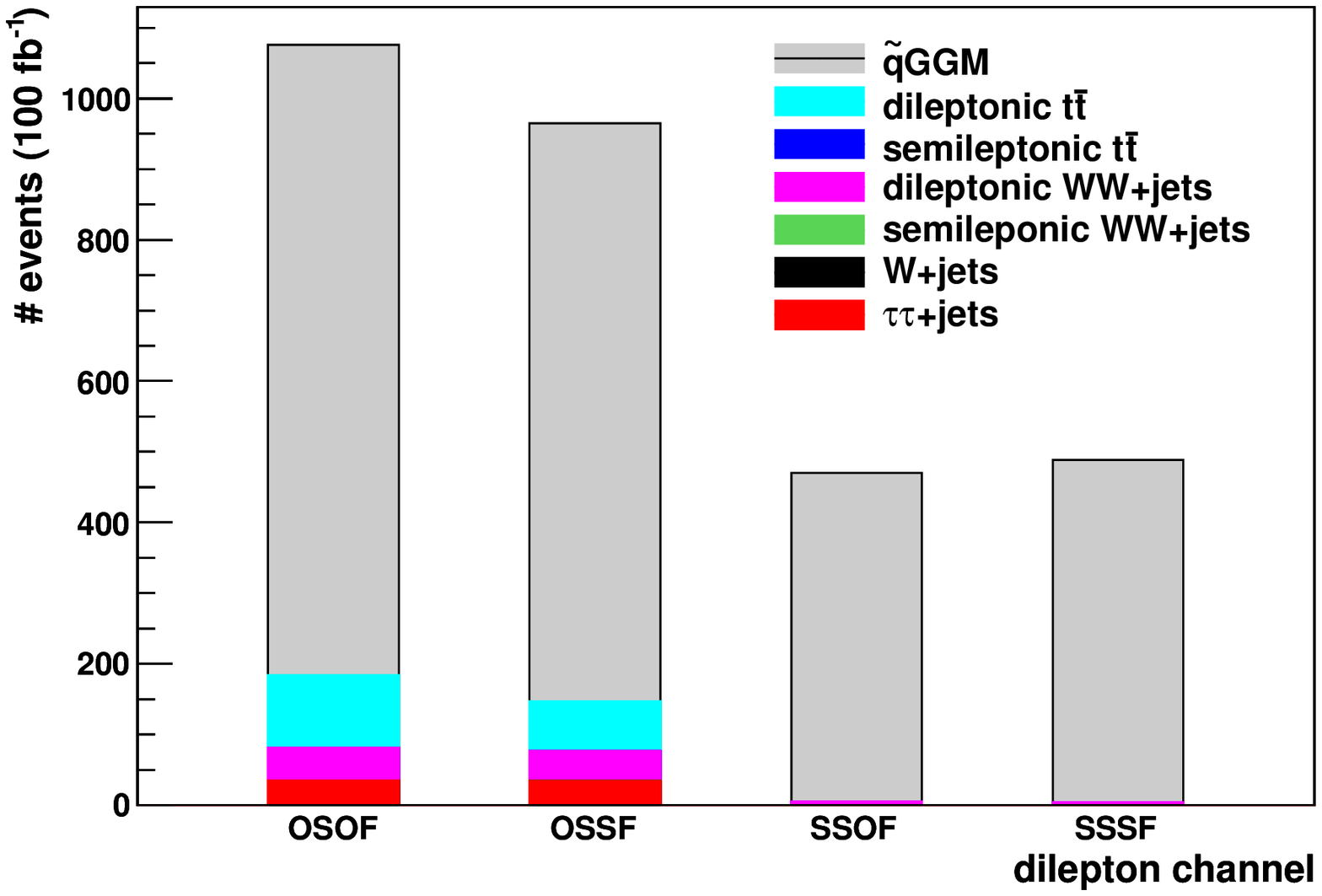}
\epsfxsize=0.49\textwidth\epsfbox{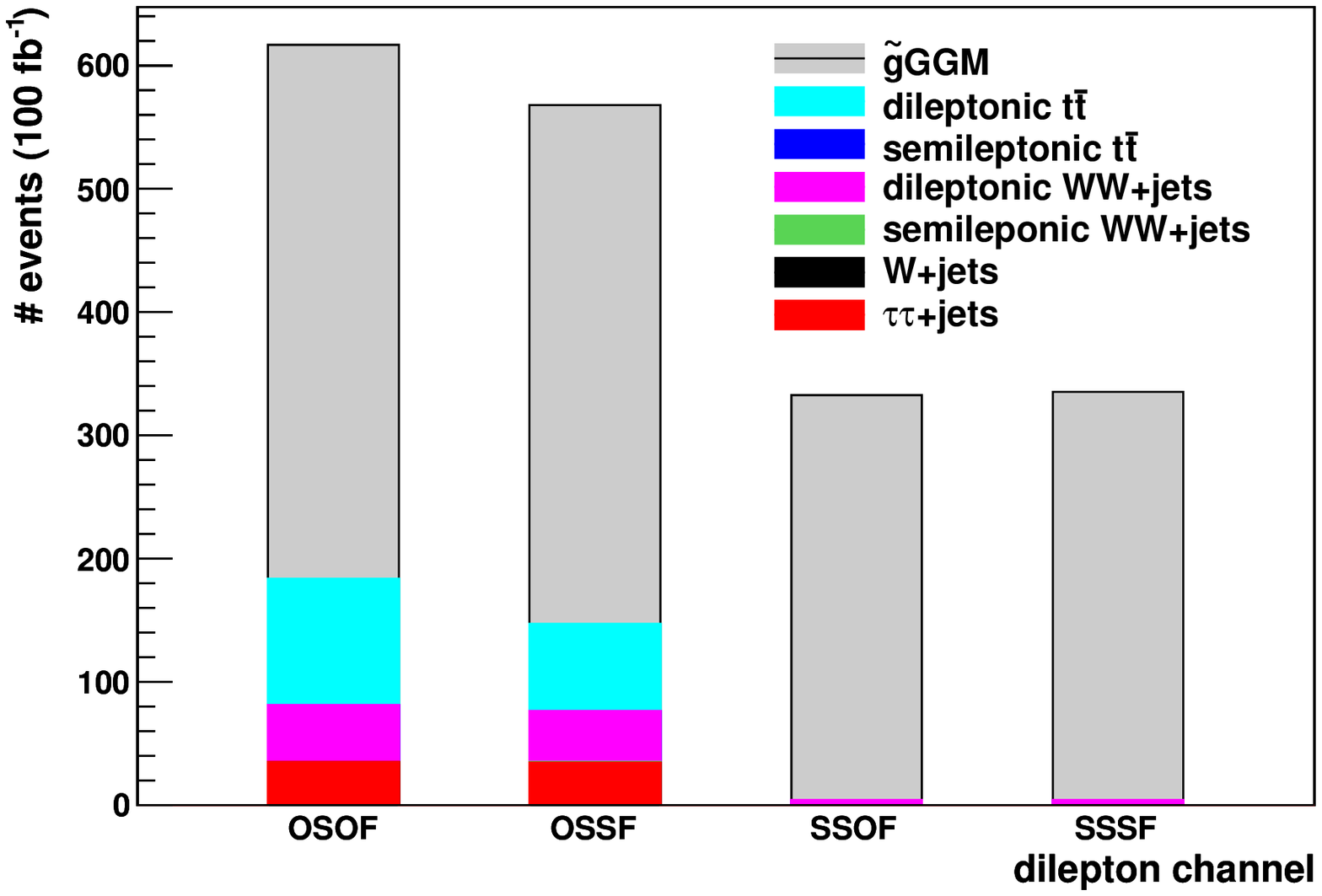}
\caption{Dilepton channels for the $\squark$GGM spectrum (left panel)
and $\gluino$GGM spectrum (right panel).  (The histograms are stacked.)  Note that OSSF is slightly
reduced due to the $Z$ veto.}
\label{fig:llCode}
\end{center}
\end{figure}

Fig.~\ref{fig:nl} shows the lepton counting at 100 fb$^{-1}$, 14 TeV, after application
of our analysis cuts.  Fig.~\ref{fig:llCode} shows a more refined view of the
 individual channels in the dileptonic events.
We see that the ratio of monoleptonic:dileptonic is of order 3:1, which
is close to what we expect from the naive counting presented in section~\ref{cascades}. 

Backgrounds, dominated by $t \bar t$+jets\footnote{We parenthetically note that in 
most of the events passing cuts,
one or both of the two leading jets typically do not come from the top's decay.}, are clearly 
not obscuring the $\squark$GGM signal.  They are more
important for $\gluino$GGM, though the dilepton mass shape is quite
different.  The background can of course be further reduced with more aggressive or
more tailored cuts, 
but we do not investigate this explicitly.

\begin{figure}[t]
\begin{center}
\epsfxsize=0.49\textwidth\epsfbox{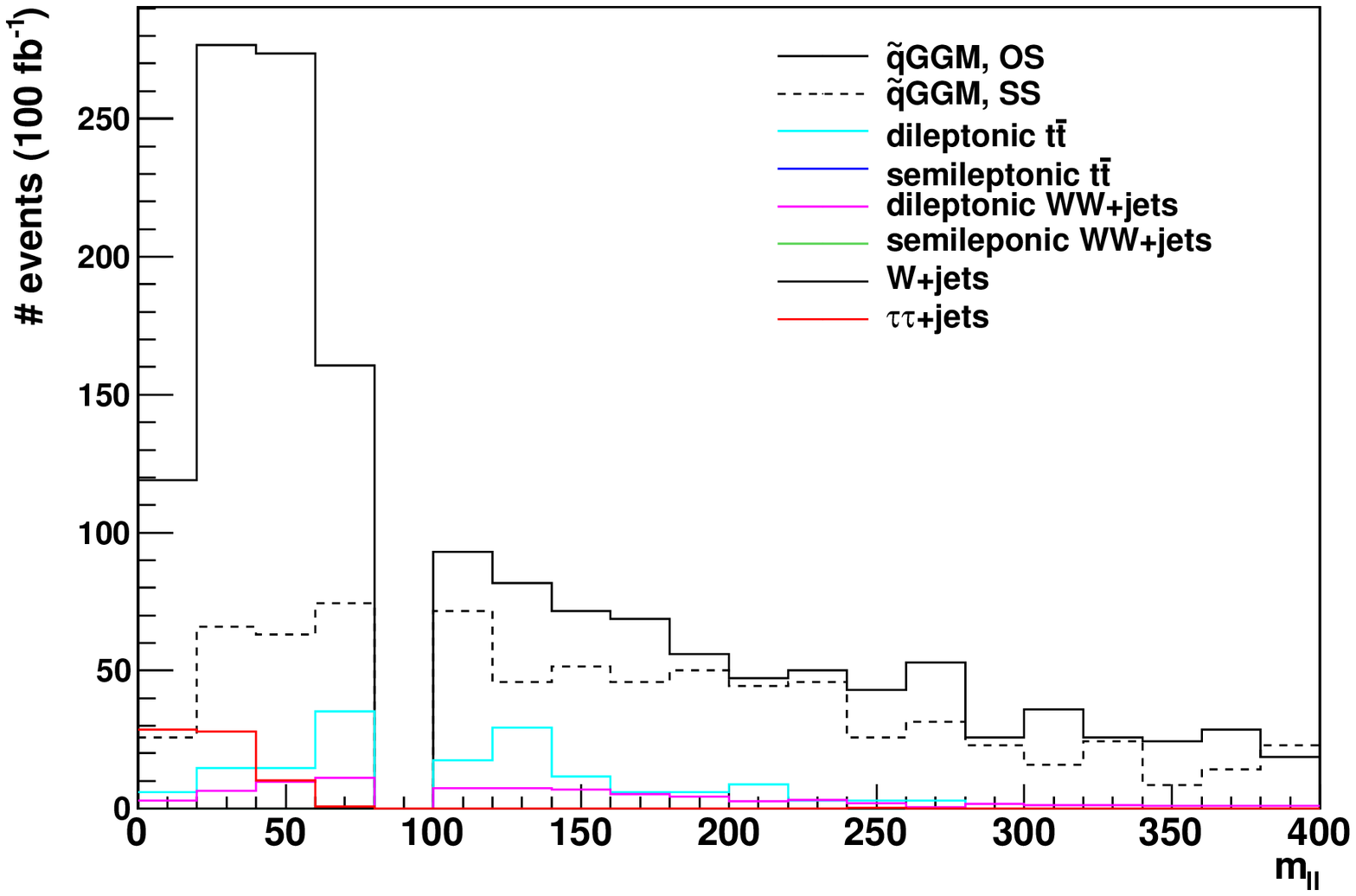}
\epsfxsize=0.49\textwidth\epsfbox{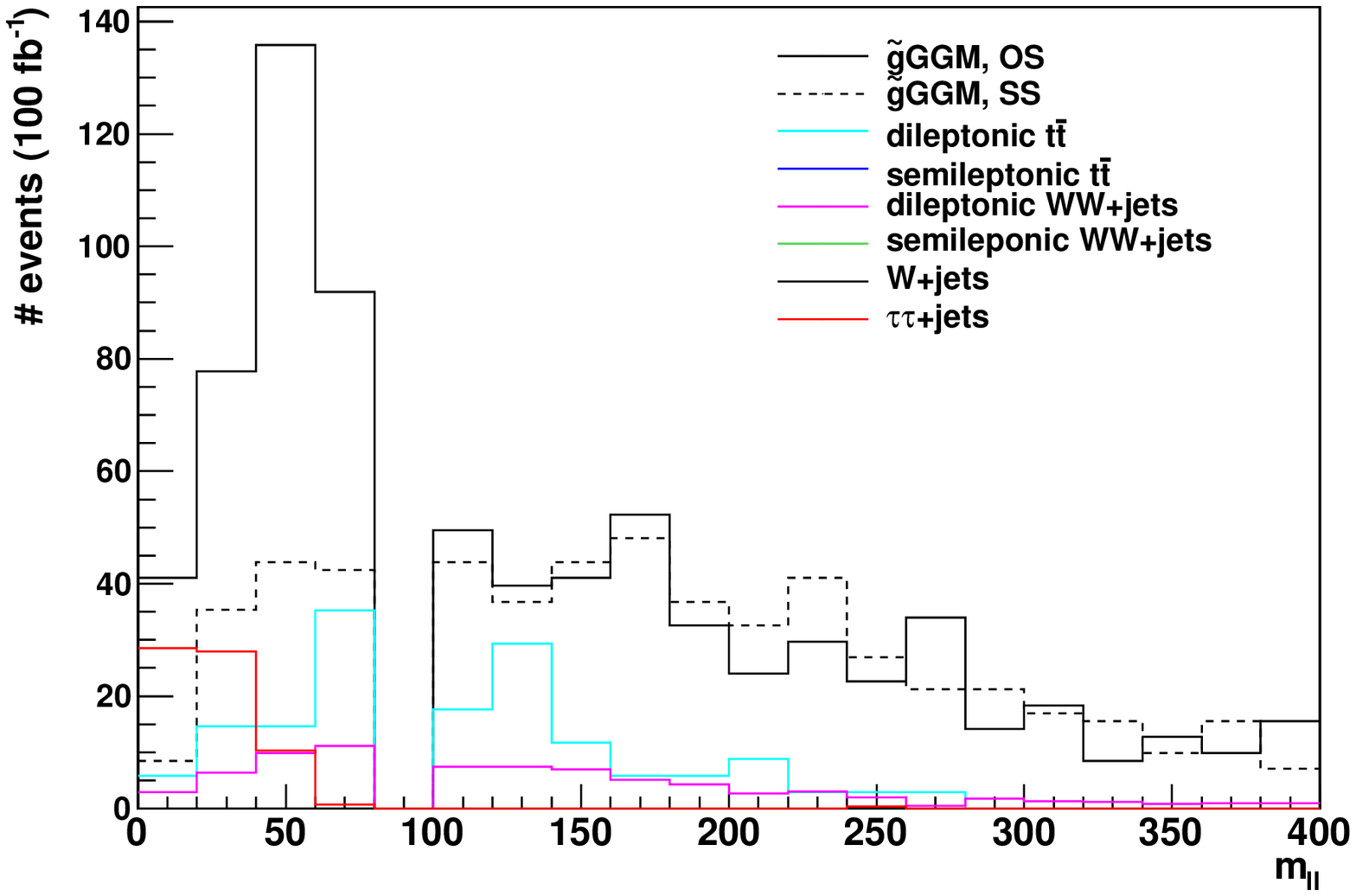}
\caption{Dilepton invariant mass distribution for OS and SS categories, for the $\squark$GGM spectrum (left panel)
and $\gluino$GGM spectrum (right panel).  The $Z$ mass window has been blinded.  (Backgrounds
are almost purely OS.)}
\label{fig:OSandSS}
\end{center}
\end{figure}

\begin{figure}[t]
\begin{center}
\epsfxsize=0.49\textwidth\epsfbox{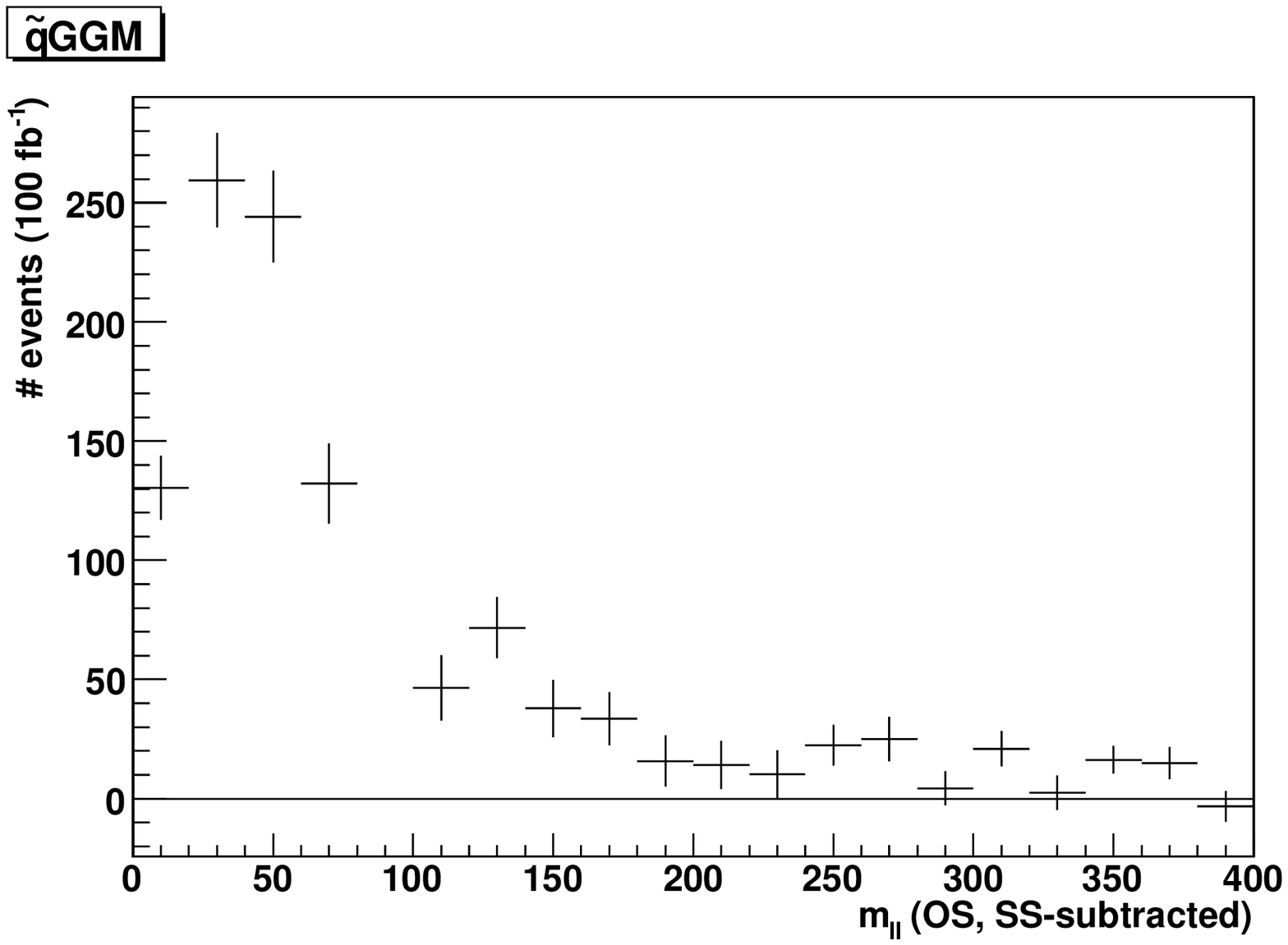}
\epsfxsize=0.49\textwidth\epsfbox{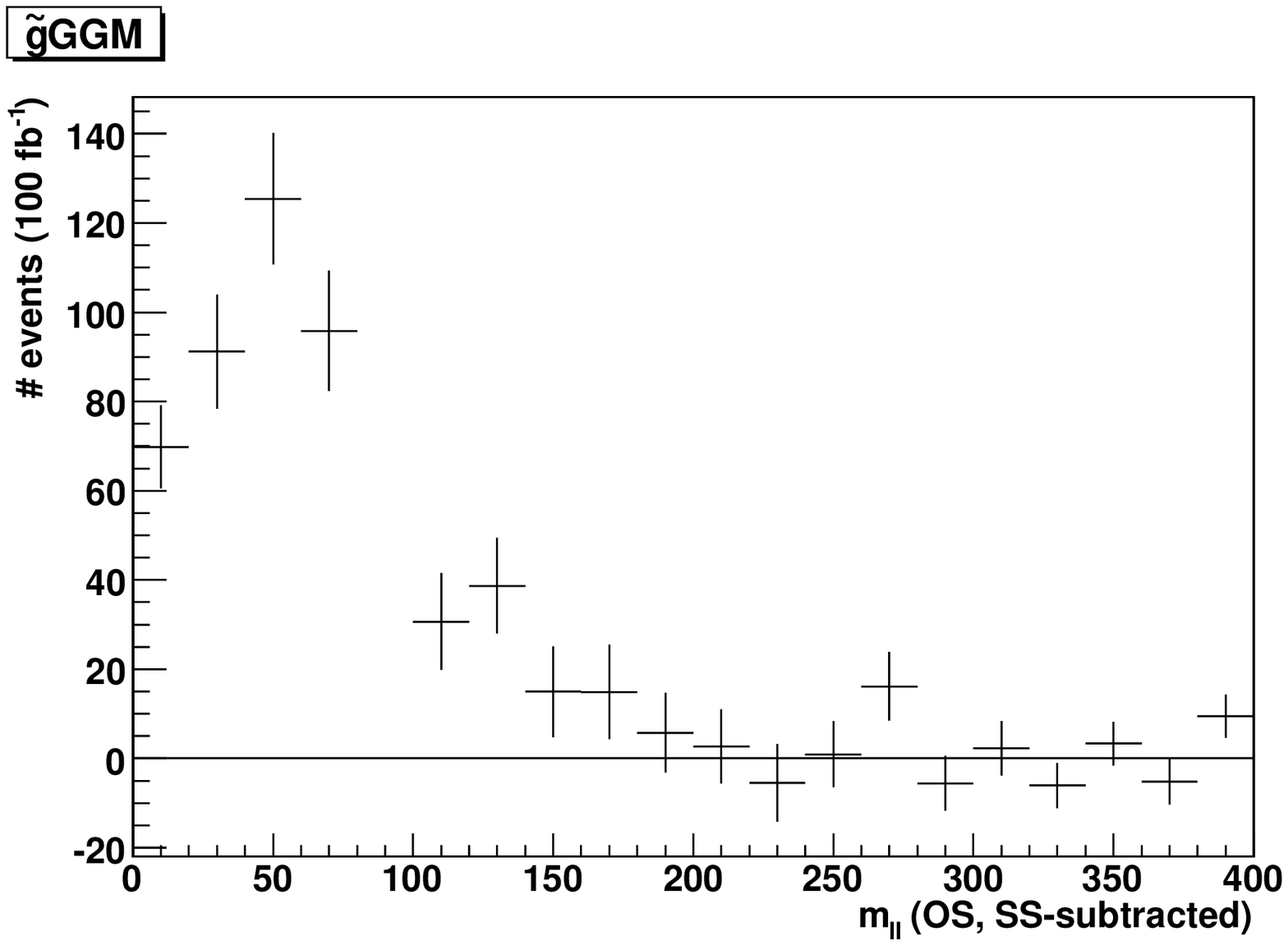}
\caption{Dilepton invariant mass distribution, applying the OS minus SS subtraction 
(including backgrounds), for the $\squark$GGM spectrum (left panel)
and $\gluino$GGM spectrum (right panel).  The $Z$ mass window has been blinded.
Error bars are representative of 100 fb$^{-1}$ statistics.}
\label{fig:subtracted}
\end{center}
\end{figure}

In Fig.~\ref{fig:OSandSS}, we show the dilepton invariant mass spectrum in the OS and SS
channels.  Fig.~\ref{fig:subtracted} displays the OS minus SS subtraction, which picks
out the OS excess.  In both plots, the $Z$ mass window 80 GeV $< m_{ll} <$ 100 GeV is blinded. 
Note that we could avoid this if we separately considered opposite-flavor channels, which are
free of $Z$ contamination.
 
In the spectrum 
$\tilde q$GGM we expect to see two bumps of comparable size, corresponding to wino and bino, sitting one
on top of the other.  Using equation~\eqref{eq:mll}, we predict peaks at 31 and 40 
GeV. This is quite consistent with what we see on the sign-subtracted invariant mass plot.
In $\tilde g$GGM spectrum, most of the decays proceed through wino, rather than bino.  Hence 
we expect to see only one bump around 50 GeV, and this is indeed what is observed.  

\begin{figure}[t]
\begin{center}
\epsfxsize=0.49\textwidth\epsfbox{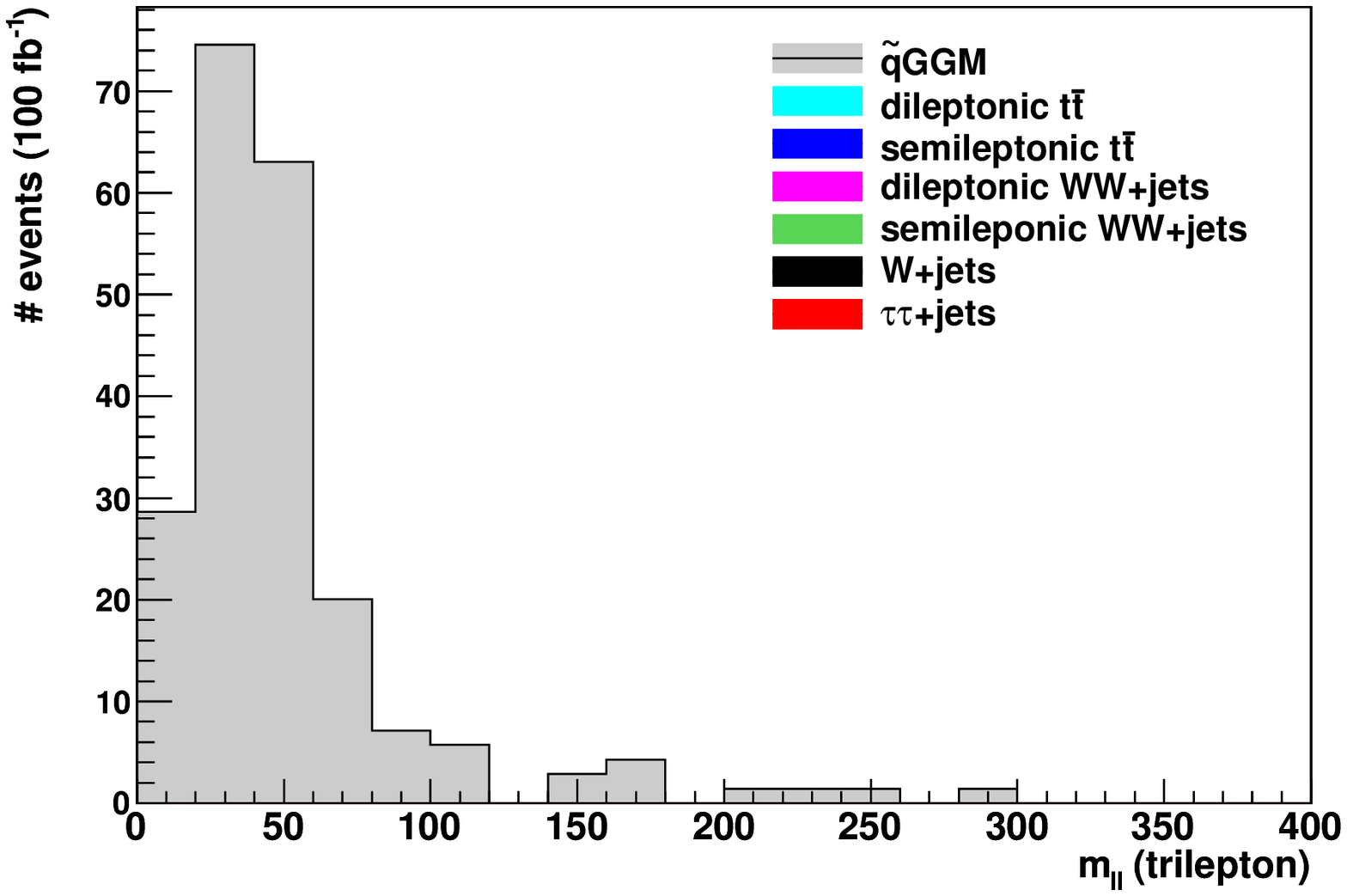}
\epsfxsize=0.49\textwidth\epsfbox{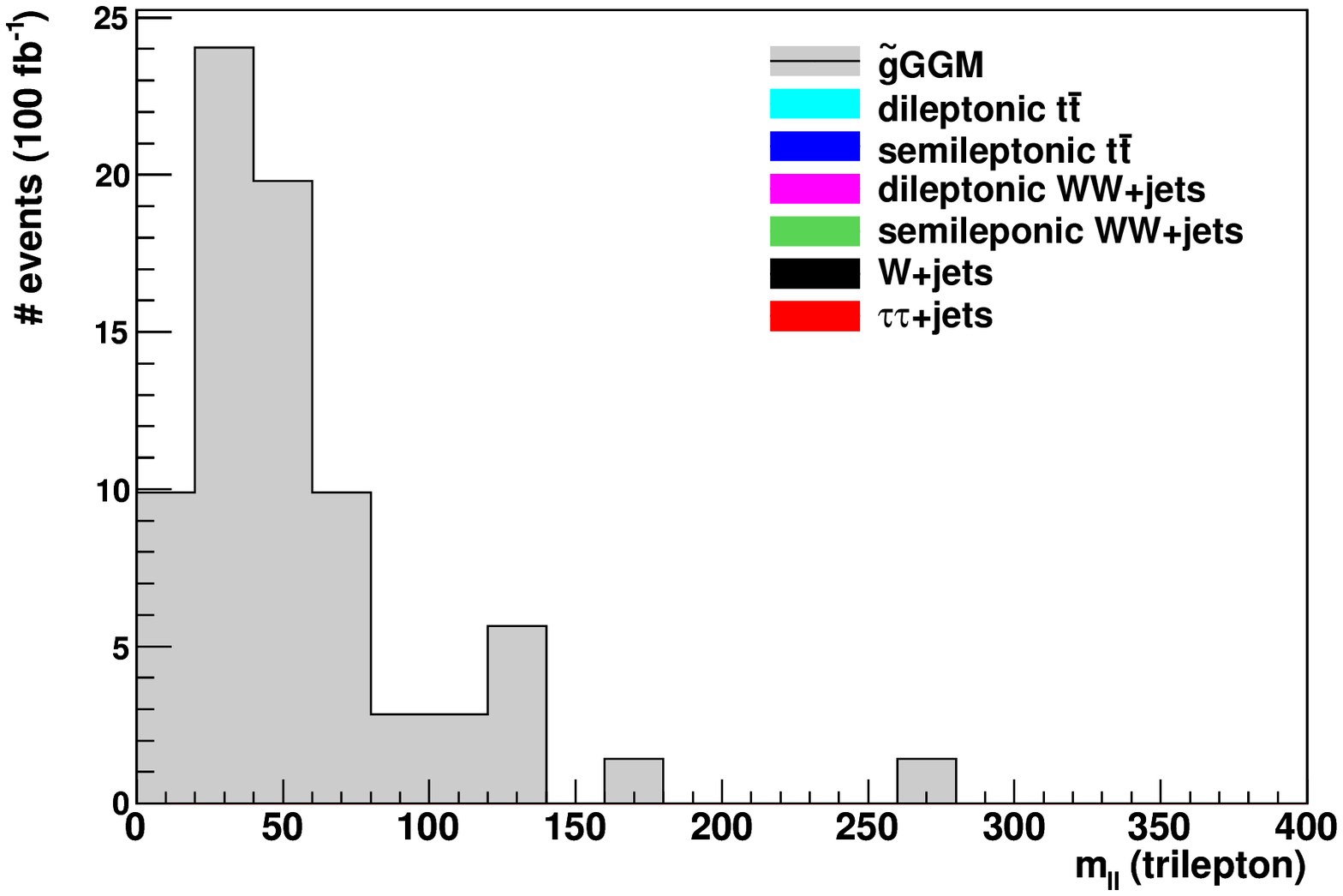}
\caption{Trilepton invariant mass distribution, obtained using the technique
described in subsection~\ref{trilep}, for the $\squark$GGM spectrum (left panel)
and $\gluino$GGM spectrum (right panel).}
\label{fig:trilepton}
\end{center}
\end{figure}

We turn now to the dileptonic invariant mass in the trileptonic channel, forming lepton pairs
according to the procedure discussed in subsection~\ref{trilep}. We show this distribution
in Fig.~\ref{fig:trilepton}. This channel essentially lacks 
any backgrounds after the cuts we impose, and reproduces the same features as the dileptonic
sign-subtracted invariant mass. 

\begin{figure}[t]
\begin{center}
\epsfxsize=0.49\textwidth\epsfbox{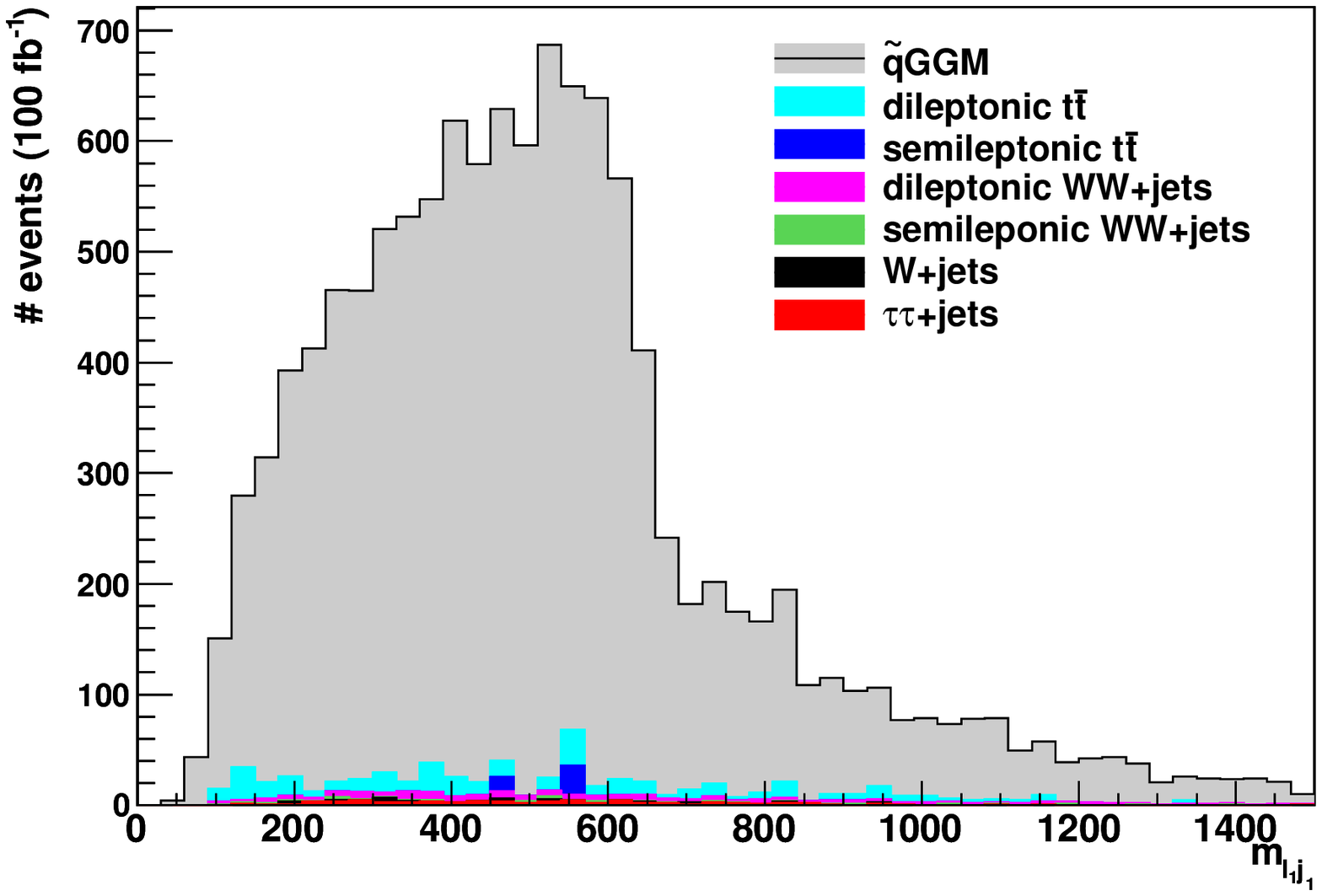}
\epsfxsize=0.49\textwidth\epsfbox{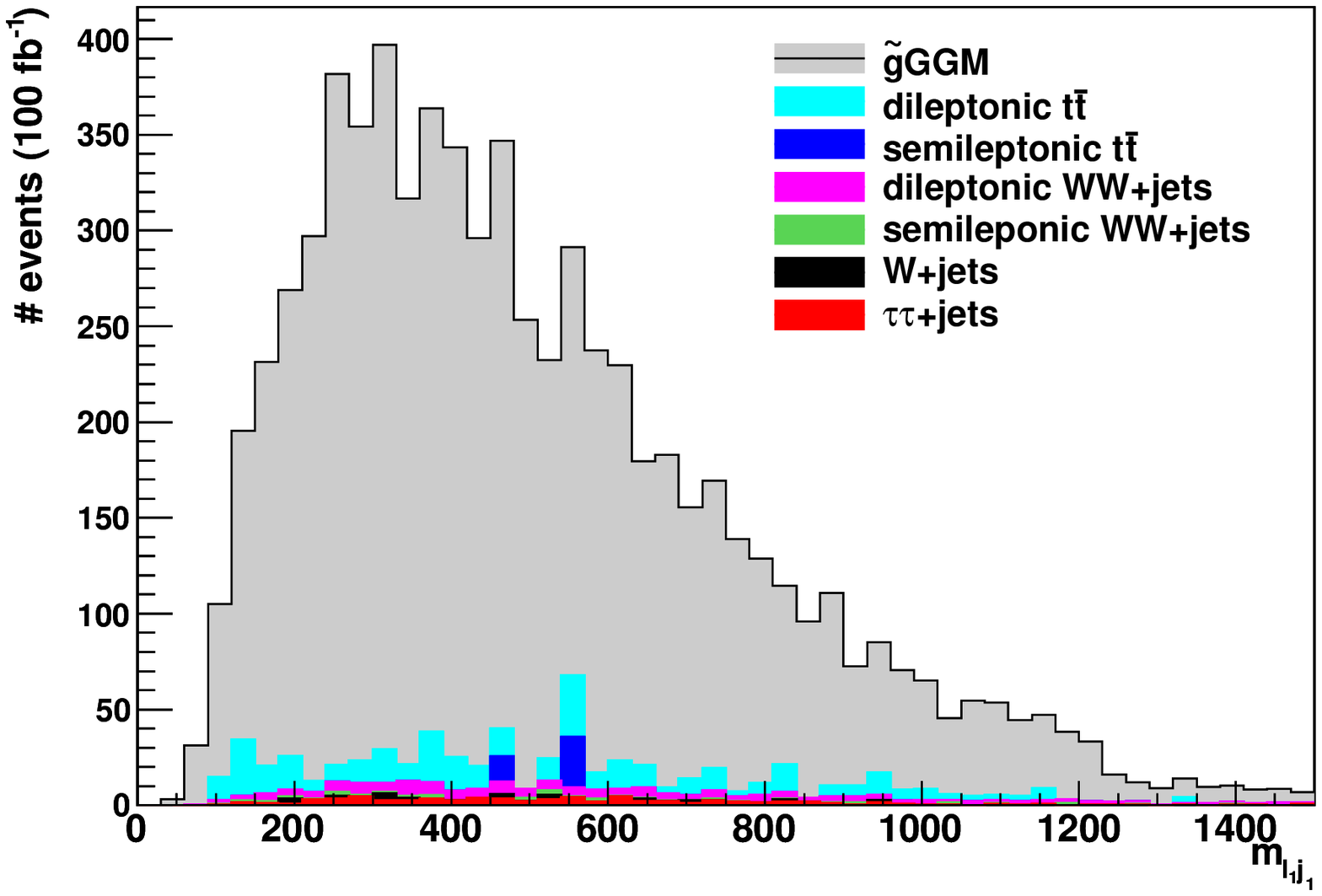}
\caption{Invariant mass of leading lepton and leading jet for the $\squark$GGM spectrum (left panel)
and $\gluino$GGM spectrum (right panel).  (The histograms are stacked.)}
\label{fig:ml1j1}
\end{center}
\end{figure}

Finally we analyze the leading-jet/leading-lepton invariant mass in Fig~\ref{fig:ml1j1}. 
Since in the $\tilde g$GGM spectrum, 
gluinos decay through the squarks off-shell, we do not see any clear 
feature.  However, in the $\tilde q$GGM spectrum we see an
edge in the distribution, corresponding to the cascade decay 
$\tilde q \to \tilde \chi \to \slep/\snu$. The 
theoretically predicted endpoint is around 625 GeV, and is easily visible.

\section{Conclusions and Outlook} \label{conc}

In this paper we began investigating the collider signals of
the largely unexplored region of MSSM parameter space with NLSP sneutrinos.  
We showed that a large portion of this region
has distinctive collider signatures at the LHC.  In particular, we focused on strong
production modes for
spectra with $\mathcal{O}$(TeV) colored superpartners, approximate flavor degeneracy in 
the LH sleptons, and RH sleptons mostly bypassed in the decay chains.

We found that these spectra lead to interesting multilepton signals.  A large
fraction of SUSY events are monoleptonic or dileptonic.  The dileptons
are mostly characterized by a broad distribution with no sign or
flavor correlations, but they are accompanied by a sizable excess of 
sign-correlated dileptons.  Unlike many SUSY dilepton signals, this excess is completely
flavor-universal.  It has a unique shape which contains information about the
mass splittings within the spectrum.
In addition, the trilepton channel has an appreciable rate.  If analyzed carefully,
it can provide strong confirmation of the physics inferred from the dileptons.
All together, these leptonic signatures are quite difficult to fake within alternative
spectra in the MSSM, and their observation should be taken as highly suggestive
evidence for a sneutrino NLSP.

We also proposed specific ways to analyze these signals. In particular, we used  
simulations of two representative spectra to demonstrate that one can extract the signal in the 
dileptonic channel using sign subtraction.  The signature in the trileptonic 
channel can also be purified, and combinatorial background 
significantly reduced, by choosing events with a unique opposite-sign pairing
between the softest lepton and one of the two hardest leptons.  This technique
works because the softest lepton is usually the one emitted from the $W^*$ in 
the decay of the slepton down to the sneutrino.

Of course, we could not cover all possible MSSM spectra with $\snu$-NLSP.
In particular, we did not analyze spectra where the RH slepton
plays an active role in the decay chains.  Though these certainly share some
common features with the spectra analyzed here, their collider signatures
will be much more ``leptogenic,'' producing up to three or four leptons
\emph{from a single decay chain}.  Indeed, spectra with active RH sleptons
can be treated as close relatives of the spectra studied in~\cite{DeSimone:2008gm, DeSimone:2009ws},
with the roles of the LH and RH sleptons interchanged.  
We will study spectra with RH sleptons in detail in a forthcoming paper~\cite{active}. 

In this paper, we tried to address only a very broad question, namely whether 
we can identify spectra with $\snu$-NLSP utilizing some well-defined collider signals. 
Taking our results as evidence that this is possible (at least in a large portion of the
allowed parameter space), we face further interesting questions.
For example, if these signals are actually discovered, then are there any additional
clues that tell us whether we are observing a high-scale or low-scale mediation scenario?
One way to answer this question might be to study flavor non-degeneracy in the sleptons
and sneutrinos.  We have avoided explicit analysis of $\tau$s, but dedicated
study of their production in SUSY events could provide useful information.  It would
be very interesting to see how feasible such a study might be at the LHC.

It would also be interesting to extend these studies to more remote parts of the parameter space.
For example, we might consider spectra where a neutralino resides between the slepton and 
sneutrino. It is also very important to understand the 
current experimental bounds on all these scenarios from LEP and the Tevatron. To the best of our knowledge, these 
studies have not been performed yet.

\acknowledgments{ We are grateful to
Kaustubh Agashe, Zacharia Chacko, Sarah Eno, Beate Heinemann, David E. Kaplan, Kirill Melnikov, 
Yasunori Nomura, David Shih, Matt Strassler, Raman Sundrum, Scott Thomas, 
Lian-Tao Wang, and especially Matt Reece for discussions and help with the {\tt BRIDGE} program.
We also thank the organizers of the workshop
``Shedding Light on Dark Matter'' at the University of Maryland, where this work was initiated.
A.K. is partially supported by NSF under grant PHY-0801323.  B.T. is supported by the Leon
Madansky Fellowship and Johns Hopkins University grant \#80020033.



\appendix

\section{Analytic approximations to opposite-sign dilepton mass distributions}
\label{mlldistrib}

The dilepton invariant mass spectrum from the chain 
$\gaugino^0 \to l_i^- \slep_i^+ \to l_i^-(l_j^+ \nu_j \snu^*_i)$
asymptotes to a polynomial in the limit where the sneutrino goes nonrelativistic
in the slepton's rest frame (or equivalently, when $m_{\slep} - m_{\snu} \ll m_{\slep}$).
Assuming constant matrix element, the normalized distribution takes the form
\beq
\frac{dP}{dm_{ll}} = \frac{6}{m_{max}^6}~m_{ll}
\left( m_{max}^2 - m_{ll}^2  \right)^2,
\eeq
where $m_{max}$ is as in equation~\eqref{eq:mll}.  This distribution peaks at
$m_{max}/\sqrt{5}$.  Putting in the electroweak current-current matrix element
for the slepton decay, this becomes
\beq
\frac{dP}{dm_{ll}} = \frac{5}{m_{max}^{10}}~m_{ll}
\left( m_{max}^2 - m_{ll}^2  \right)^3 \left( m_{max}^2 + 3m_{ll}^2 \right),
\eeq
which peaks at 
\begin{displaymath}
\sqrt{ \frac{1+\sqrt{28}}{27} } m_{max} \simeq (0.48)m_{max}~.
\end{displaymath}

\section{Cross sections}\label{tables}
In this appendix, we summarize leading-order production cross sections for our signals 
and backgrounds in section~\ref{signat}. 

The following table shows the cross sections (in fb) for the different super-QCD
pair production modes for our two sample spectra.
\begin{center} 
\begin{tabular}{|c|c|c|c|c|}\hline
spectrum & $\tilde g \tilde g$ & $\tilde g \tilde q$ & $\tilde q \tilde q^*$ & $\tilde q \tilde q$\\
\hline 
$\gluino$GGM & 650 & 310 & 8 & 36\\
$\tilde q$GGM & 10 & 240 & 250 & 440 \\
\hline
\end{tabular}
\end{center} 
\vspace{0.7cm}

The next table lists the inclusive super-QCD pair production cross sections (in fb) for our sample spectra, 
broken down into multilepton channels, before and after reconstruction and cuts.
\begin{center} 
\begin{tabular}{|c||c|c|c|c|}\hline 
 & \multicolumn{2}{|c|}{  $\tilde q$GGM } & \multicolumn{2}{|c|}{ $\tilde g$GGM } \\ \hline
 & partonic & reco & partonic & reco \\ \hline \hline
total & 950 & 310 & 1000 & 148 \\  \hline
0l & 450 & 200 & 350 & 85 \\
1l & 350 & 80 & 450 & 48 \\
2l & 120 & 26 & 190 & 15 \\
3l & 19 & 3.6 & 27 & 1.4 \\
4l & 1.9 & 0.3 & 1.7 & 0.04 \\ \hline 
\end{tabular}
\end{center} 
\vspace{0.7cm}

The last table lists our major backgrounds, including cross-sections at generator-level and
in the various multi-lepton channels after reconstruction.
The matched \ttbar+jets sample is inclusive, while rest of the backgrounds were 
produced with generator-level cuts, as described in subsection~\ref{backgrounds}.
When we observed no events after cuts, we placed a ``0'' in the entry, but this
is roughly to be understood as a limit of $\simlt$ 1 event after a 100 fb$^{-1}$ run (i.e.,
cross section less than about 0.01 fb).  
\begin{center} 
\begin{tabular}{|l||c|c||c|c||c|c||c|c|} \hline 
                 &  dilep \ttbar+jets & semilep \ttbar+jets & dilep $WWjj$ & semilep $WWjj$ & $Wjj \to (l/\tau)\nu jj$ & $\tau^+\tau^-jj$  \\
\hline \hline
    generated    &    6.3$\times10^4$ &     2.5$\times10^5$ &          104 &            418 &                     4000 &             4400  \\ \hline
   1l reco       &                2.1 &                0.40 &         0.60 &     $\sim$0.02 &                     0.31 &                0  \\ 
   2l (OS) reco  &                1.7 &                   0 &         0.88 &              0 &                        0 &             0.75  \\       
   2l (SS) reco  &                  0 &                   0 &         0.09 &              0 &                        0 &                0  \\
   3l reco       &         $\sim$0.03 &                   0 &            0 &              0 &                        0 &      $\sim$0.008  \\ \hline
\end{tabular}
\end{center} 
\vspace{0.7cm}

\bibliography{lit}
\bibliographystyle{apsper}
\end{document}